\documentclass[twoside,leqno]{article}

\usepackage[letterpaper]{geometry}

\usepackage{siamproceedingsArxiv}

\usepackage[T1]{fontenc}
\usepackage{amsfonts}
\usepackage{graphicx}
\usepackage{epstopdf}
\usepackage{enumitem}
\usepackage{xcolor}
\usepackage{algorithmic}
\ifpdf
  \DeclareGraphicsExtensions{.eps,.pdf,.png,.jpg}
\else
  \DeclareGraphicsExtensions{.eps}
\fi

\newsiamremark{remark}{Remark}
\newsiamremark{hypothesis}{Hypothesis}
\crefname{hypothesis}{Hypothesis}{Hypotheses}
\newsiamthm{claim}{Claim}

\usepackage{amsopn}

\usepackage{booktabs}
\usepackage{enumitem}
\usepackage{subcaption}
\newcommand{\ee}{\mathbb E}

\newcommand{\abs}[1]{\left|{#1}\right|}

\newcommand{\tth}{^\text{th}}

\newcommand{\txt}[1]{\text{#1}}

\newcommand{\stext}[1]{\ \ \ \ \ \text{(#1)}}

\newcommand{\snc}[1]{\stext{since ${#1}$}}

\newcommand{\ipncm}[3]{\begin{figure}[H]\begin{center}\includegraphics[scale = {#1}]{../Media/#2.pdf}\caption{#3}\end{center}\end{figure}}

\renewcommand{\ipncm}[3]{\begin{figure}[H]\begin{center}\includegraphics[scale = {#1}]{Media/#2.pdf}\caption{#3}\end{center}\end{figure}}
\newcommand{\cS}{\mathcal{S}_k}

\newcommand{\ind}[1]{\mathbf{1}\{#1\}}

\begin{document}

\newcommand\relatedversion{}

\title{\Large Envy-free Randomized Apportionment\relatedversion}
    \author{Francesco La Commare\thanks{Yale University (\email{francesco.lacommare@yale.edu}, \email{j.tuckerfoltz@siam.org}).}
    \and Jamie Tucker-Foltz\footnotemark[1]    
    \and Victor Verdugo\thanks{Pontificia Universidad Católica de Chile
  (\email{victor.verdugo@uc.cl}).}}

\date{}

\maketitle






\pagenumbering{arabic}
\setcounter{page}{1}

\begin{abstract}
	Nearly every form of representative democracy faces a mathematical quandary known as \emph{apportionment}. The United States allocates representatives to each state in proportion to its population, while many parliamentary democracies select representatives from party lists in proportion to the number of votes each party receives. In both settings, the legislature's small size necessitates rounding to the nearest whole number of seats. There is no canonical way to determine who gets rounded up and who gets rounded down, with classic impossibility results holding that any deterministic apportionment system suffers from paradoxes.
	
	Recent work has shown a plethora of ways to randomly round seat shares, in which the various rounding events are correlated to guarantee a specific total number of seats. While this approach can achieve perfect ex ante fairness, all known methods suffer from a new kind of paradox to which deterministic methods are immune: Party (state) $i$ may have a larger number of votes (population) than party $j$ but receive fewer seats. In this case, we say that party $i$ \emph{envies} party $j$. We investigate the extent to which such envy is avoidable. Specifically, we seek to design apportionment rules that minimize the expected number of envious pairs $(i, j)$. If the expectation is zero, we say that the seat allocation is \emph{envy-free}.
	
	On the negative side, we show that any $n$-party apportionment rule that ignores lower quotas (the floors of the deserved number of seats for each party) suffers from a worst-case envy of $\Theta(n^2)$, even on instances where an envy-free allocation exists. All previously studied apportionment rules fall into this class and thus do a poor job of minimizing envy, achieving the worst possible multiplicative and additive approximations (up to constant factors). In the case where all lower quotas are equal, we prove that all allocations have the same expected envy. In the general case, we give polynomial-time algorithms to decide whether an envy-free allocation exists and, if not, to construct an envy-minimizing apportionment rule. We also consider a handful of alternative objectives, such as minimizing total envy variance. We introduce a general LP framework giving polynomial-time algorithms for a large class of such objective functions by efficiently solving the separation problem on the dual. Finally, we empirically demonstrate that envy-free allocations exist very frequently, including in every historical instance from the US decennial census. Taken together, these results provide a clear recipe for implementing randomized apportionment in a less objectionable way in practice.
\end{abstract}

\section{Introduction}\label{secIntro}

What is the best way to round a list of rational numbers to integers while preserving the sum? In 1921, this seemingly innocuous question caused a constitutional crisis in the United States. The list in question was
\begin{align*}
	(&9.70, 1.38, 7.24, 14.16, 3.88, 5.70, 0.92, 4.00, 11.97, 1.78, 26.8, 12.11, 9.93, 7.31, 9.99, 7.43, 3.17, 5.99,\\&15.92, 15.16, 9.86, 7.40, 14.07, 2.27, 5.36, 0.32, 1.83, 13.04, 1.49, 42.91, 10.57, 2.67, 23.8, 8.38, 3.24,\\&36.03, 2.50, 6.96, 2.63, 9.66, 19.27, 1.86, 1.46, 9.54, 5.61, 6.05, 10.88, 0.80).
\end{align*}
These represented the numbers of seats in the US House of Representatives that each state deserved based on its population from the 1920 census, also known as \emph{standard quotas}. The constitution requires House seats to be distributed in proportion to these numbers, but supplies no formula for deciding how to round them. For instance, the largest number, 42.91, is the standard quota for the most populous state of New York, calculated as
$$\frac{(\txt{state population})}{(\txt{total US population})} \cdot (\txt{house size}) = \frac{10385228}{105273051} \cdot 435 \approx 42.91.$$
Should it be rounded up to 43, down to 42, or perhaps to something else entirely? In 1921, congress was unable to agree on a coherent rounding policy for all 48 US states, and thus violated their constitutional duty to reapportion seats! The numbers from the previous 1911 apportionment remained in place until 1929, by which point they were woefully unfair and out of date~\cite{BadNoApportionment}.

There is no obvious solution to this rounding problem. Simple policies like rounding to the nearest integer do not result in a fixed house size. Thomas Jefferson proposed a modified rounding scheme, used in the US for several decades, in which standard quotas are scaled up by a constant factor $\lambda$, then rounded down to integers, with $\lambda$ carefully chosen to ensure the correct house size. John Adams proposed the opposite: scale down by a constant factor, then round up. These methods are \emph{not} equivalent---Jefferson's method gives New York 45 seats whereas Adams' method gives only 41. And paradoxically, neither of these numbers is within one of the standard quota of 42.91.

In the absence of a canonical rule, the early history of the United States was fraught with decennial political battles over the choice of apportionment method. Today, the US uses an algorithm devised by mathematicians, building on the ``divisor-based'' approach of Jefferson's and Adams' methods. Outside the US, apportionment is used around the world in parliamentary systems, where parties are awarded seats in proportion to the number of votes cast for each party. For a comprehensive survey of the various apportionment systems in place around the world and their theoretical underpinnings, we refer the reader to the book by Pukelsheim~\cite{Pukelsheim17}.

The first modern, mathematical investigation of apportionment was initiated in the 1970s in a series of papers by Balinski and Young~\cite{BY74,BY75,BY77,BY79,BY79a}. They applied an axiomatic approach to the study of apportionment methods, identifying desirable properties and proving/disproving the existence of rules satisfying them. Among the most famous results is that any rule satisfying the \emph{quota axiom}---which requires the number of seats awarded to a state be either the ceiling or floor of its standard quota---suffers from the \emph{population paradox}: it is possible for one state to rise in population and lose a seat, while another state falls in population and gains a seat~\cite{BY01}.

In 2004, Grimmett~\cite{Grimmett} proposed a radical new manner to apportion seats: by random lot. In the foregoing example, New York would be granted 42 seats with probability 9\% and 43 seats with probability 91\%. Not only does this avoid the population paradox (in expectation), but it also ameliorates the inherent unfairness associated with deterministic apportionment methods. Each party/state receives the number of seats it deserves exactly ex ante, and up to an additive error of at most one seat ex post. Furthermore, by correlating these lotteries, one can ensure that exactly the right number are rounded up, so the house size always works out to the correct value.

Grimmett's original construction of such correlated lotteries was rather ad hoc, and subsequent work on randomized apportionment has sought to identify more principled approaches with stronger axiomatic properties. We discuss these developments in Section~\ref{subLitReview}. However, none of them confront what is arguably the biggest elephant in the room, possibly the largest barrier to adoption. One state may have a higher population but be apportioned fewer seats than another. Or in the parliamentary context, one party $i$ may receive more votes than party $j$ but fewer seats. (We will adopt the language of parties and votes rather than states and populations going forward.) When this happens, we say that party $i$ \emph{envies} party $j$. The fundamental questions we ask in this paper are: \emph{When} and \emph{how} envy is avoidable?

\subsection{Example}\label{subExample}

The following minimal example illustrates how envy arises, and also how it may be avoided. Consider a nation with 100 voters, $H = 10$ seats in the legislature, and $n = 4$ parties, numbered 1, 2, 3, 4. Suppose the number of votes for each party are respectively 14, 18, 32, 36. As previously noted, the first step is to calculate the standard quota of each party, which is the number of seats they deserve based on scaling the raw vote tallies down to the size of the legislature. The integral part of the standard quota is called the \emph{lower quota} and the fractional part the \emph{residue}, which are tabulated in Table~\ref{tabIntroExample}.
\begin{table}[h!]
	\centering
	\begin{tabular}{|c|c|c|c|c|} 
		\hline
		Party & Num votes & Standard quota & Lower quota & Residue\\
		\hline
		1 & 14 & 1.4 & 1 & 0.4\\
		2 & 18 & 1.8 & 1 & 0.8\\
		3 & 32 & 3.2 & 3 & 0.2\\
		4 & 36 & 3.6 & 3 & 0.6\\
		\hline
	\end{tabular}
	\caption{An instance of the apportionment problem with $n = 4$ and $H = 10$.}
	\label{tabIntroExample}
\end{table}
\vspace{-.4cm}

Throughout this paper, we restrict ourselves to the class of quota-satisfying and proportional apportionment rules. In other words, each party must be granted a number of seats equal to at least its lower quota, with some parties receiving one extra seat, where an extra seat is awarded with probability equal to the party's residue. The first such rule explicitly proposed in the context of apportionment is \emph{Grimmett's method} \cite{Grimmett}. It arranges the residues as consecutive blocks on a number line, each of length equal to the corresponding residue. Then we imagine placing a comb with teeth separated by width 1 on top of these blocks as in Figure~\ref{figIntroExampleBadOrder}, with an offset drawn uniformly from $[0, 1]$. The teeth determine which parties get extra seats.

\ipncm{.44}{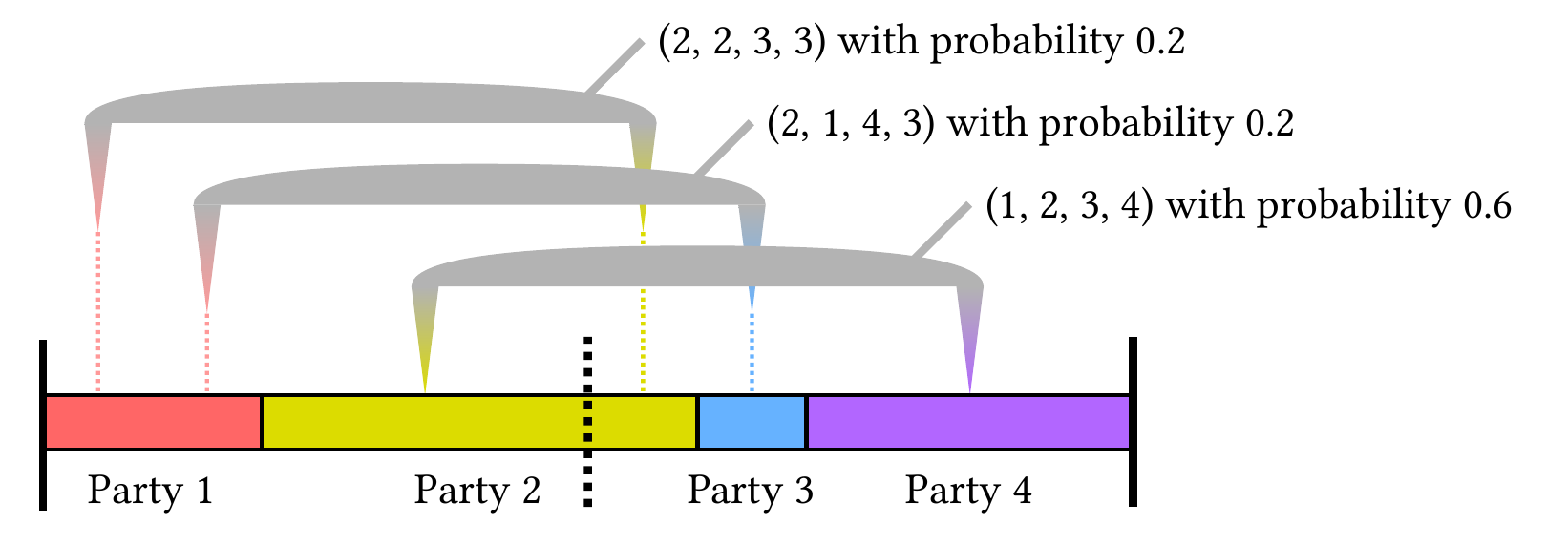}{\label{figIntroExampleBadOrder}The outcome of running Grimmett's method using the order of parties given: 1, 2, 3, 4.}

In this case, there are three possible outcomes, one of which is quite undesirable. Specifically, there is a 20\% probability that the number of seats for parties 1, 2, 3, and 4 will be respectively $(2, 1, 4, 3)$. This outcome leaves both parties 2 and 4 envious: Party 1 received fewer votes but more seats than party 2, while party 3 received fewer votes but more seats than party 4.

Grimmett's method has been criticized before on the grounds that the order is arbitrary, and thus it is not entirely unexpected that situations like this can happen. Sampford rounding~\cite{Sampford} has been explicitly proposed as an alternative, due to its useful axiomatic properties~\cite{MonotoneRandomizedApportionment}. On this instance, Sampford rounding reduces, but does not entirely eliminate, both of these envious pairs. The result is

$$\txt{Num seats (Sampford)} = \begin{cases}
	(2, 1, 4, 3) & \txt{with probability } 14/505 \approx 0.03\\
	(1, 1, 4, 4) & \txt{with probability } 27/505 \approx 0.05\\
	(2, 1, 3, 4) & \txt{with probability } 60/505 \approx 0.12\\
	(1, 2, 4, 3) & \txt{with probability } 60/505 \approx 0.12\\
	(2, 2, 3, 3) & \txt{with probability } 128/505 \approx 0.25\\
	(1, 2, 3, 4) & \txt{with probability } 216/505 \approx 0.43\\
\end{cases},$$
entailing that parties 2 and 4 will each only be envious with probability $\frac{74}{505} \approx 0.15$.

\ipncm{.44}{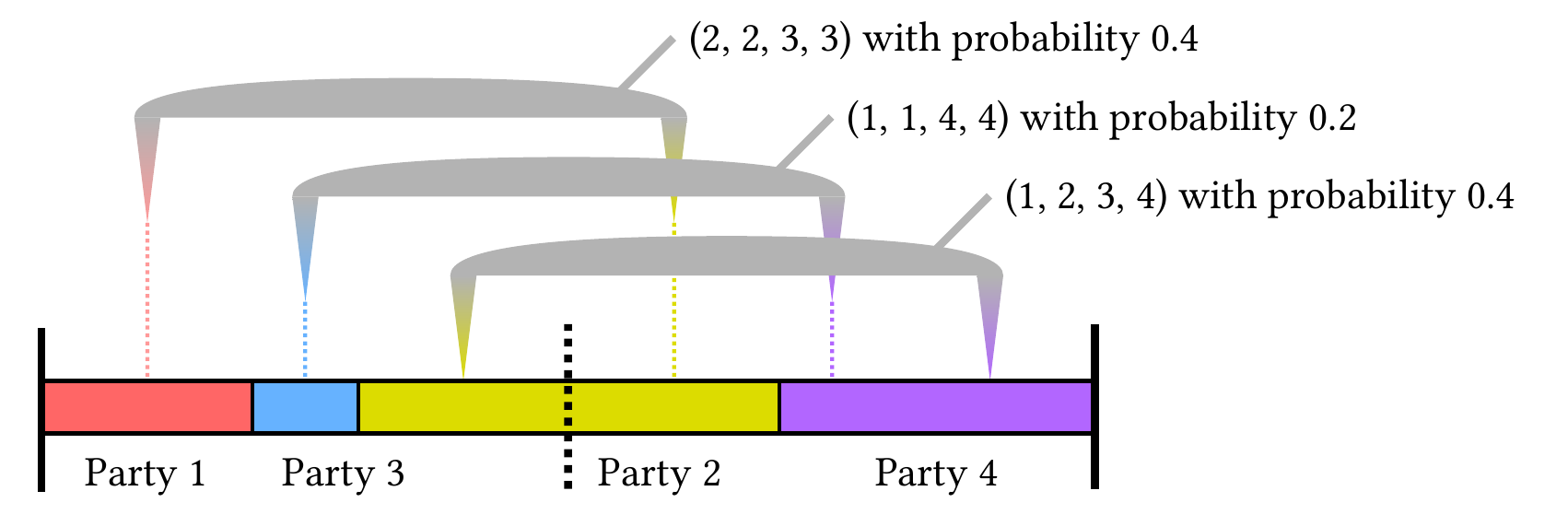}{\label{figIntroExampleGoodOrder}The outcome of running Grimmett's method using the order 1, 3, 2, 4.}

However, this is not the best we can do. If we instead run Grimmett's method using the order 1, 3, 2, 4, it turns out we can eliminate the envy entirely. As shown in Figure~\ref{figIntroExampleGoodOrder}, the randomized apportionment is supported only on allocations that are monotonically increasing in the order of parties: $(2, 2, 3, 3)$, $(1, 1, 4, 4)$, and $(1, 2, 3, 4)$. We thus say that the seat allocation is \emph{envy-free}.

The goal of this paper is to understand how much envy is inevitable and how much envy can be minimized by a clever choice of allocation.

\subsection{Our Contributions and Technical Overview}\label{subContributions}

Our results about envy in randomized apportionment include impossibilities, efficient algorithms, and empirical validation. In Section~\ref{secPrelim} we review the standard apportionment model and discuss Grimmett's and Sampford's methods which we use as baseline references throughout. We also formally introduce our fundamental definition of envy, through multiple different objective \emph{envy aggregation functions}. Throughout the paper we focus mostly on the particular objective of minimizing expected number of envious pairs of parties.

In Section~\ref{secSameFloor} we study the special case where all parties share the same lower quota. We prove that, in this setting, the total expected envy is \emph{rule-independent}: every quota-satisfying and proportional apportionment rule yields the same value, given in closed form by Equation~\eqref{eq:E-closed-form}. As a corollary, we obtain a tight upper bound of $k(n-k)/2$ when all residues are equal, where $k \leq n - 1$ is the integral sum of residues.

In Section~\ref{secMultifloor} we turn to the general multi-floor case, where rule-independence breaks down and envy minimization becomes a genuine optimization problem. We first show that \emph{oblivious} apportionment rules, those whose randomization depends only on residues and ignores lower quotas, can be $\Theta(n^2)$-suboptimal: we exhibit instances admitting an envy-free rule for which every oblivious rule incurs $\Omega(n^2)$ expected envy, matching the universal upper bound up to constant factors. This holds even for a non-adaptive adversary; i.e., a single distribution over instances thwarts any algorithm. This negative result applies to all previously studied randomized apportionment methods, including Grimmett's method and Sampford rounding.

We then present a linear-programming-based approach to study total envy. First, we introduce a compact flow-based linear program $\ref{form:EF}$ with $O(nk)$ variables and constraints that decides whether an envy-free proportional rule exists. Then, we introduce an exponential-size linear program \ref{form:minLP} that minimizes expected total envy when zero envy is not achievable. 
We show that the dual of \ref{form:minLP} can be solved in polynomial time via the ellipsoid method, using a separation oracle that reduces to a polynomial-time dynamic program.
Our LP approach can handle more general envy-aggregation functions, such as envy deviations or the probability of not being envy-free; in Section~\ref{sec:envy-agg}, we introduce these indicators, and in Section~\ref{subsecMinEnvy}, we introduce our framework for minimizing the expectation of {\it block-aggregable} functions.
As a corollary, using our LP framework, we can find in polynomial time the apportionment rule that minimizes total envy variance for single-floor instances; key to this result is our rule-independence theorem for the expected total envy, which reduces the variance minimization problem to a linear program.

In Section~\ref{secEmpirical} we complement the theory with experiments on both synthetic and real-world instances. In random instances, envy-free solutions exist with high probability once the house size $H$ is moderately larger than the number of parties $n$, and all canonical oblivious rules incur substantially more envy than the optimum. On historical U.S.\ census data, we test three preprocessing methods that accommodate the constitutional constraint of at least one seat per state, and find that an envy-free proportional rule exists for every decade under all three methods. Taken together, these results suggest that envy-free randomized apportionment is not merely a theoretical possibility but a practical one, achievable in the very instances that motivate the problem.

\subsection{Related Work}\label{subLitReview}

The problem of selecting a subset of exactly $k$ objects from a set of $n$ objects with given marginal inclusion probabilities is well studied in statistics, under the name \emph{$\pi$ps sampling} ({\it probabilities proportional to size}). We refer the reader to the survey by Brewer and Hanif~\cite{BrewerHanif} and the book by Till\'e~\cite{TilleBookAlgorithms} for a comprehensive overview. 
While Grimmet~\cite{Grimmett} was the first to apply this problem to apportionment, his algorithm is equivalent to the method of \emph{systematic sampling} developed by Madow~\cite{Madow49} more than 50 years earlier; curiously, this method has been independently re-invented multiple times, including Boix-Adser\`a et al.~\cite[Algorithm 3]{Blotto}. In the computer science literature, this problem is often referred to as \emph{dependent rounding}, which has been extensively used in the design of relax-and-round approximation algorithms~\cite{Gandhi}.

Our work contributes to a growing field of research on algorithmic apportionment and randomized apportionment specifically. G\"olz et al.~\cite{GoltzHouse} study whether it is possible to correlate lotteries across different house sizes so that passing from $H$ to $H + 1$ can be implemented by giving an extra seat to some party without taking away any seats. In other words, the apportionment must be a lottery over deterministic rules that each satisfy \emph{house monotonicity}~\cite{BY01}. Cembrano et al.~\cite{CembranoCombInsights} give a further characterization of the space of randomized rules that satisfy this axiom. Correa et al.~\cite{MonotoneRandomizedApportionment} study a different notion of monotonicity in which arbitrary subsets of parties may form coalitions, and the objective is to avoid paradoxes where a coalition could gain support but lose seats. They identify Sampford rounding as a promising contender, succeeding where many other methods fail. We show that Sampford rounding does \emph{not} perform well at minimizing envy, simply because it ignores lower quotas. Cembrano et al.~\cite{CembranoOnlineApportionment} extend the static apportionment problem to an online setting. For randomized rules, they show that for $n \leq 3$ parties, it is possible for each party, in expectation, to receive its proportional share in every step. In the realm of deterministic apportionment, there has been extensive work employing techniques similar to those we use, such as linear programming and network flows~\cite{Cembrano2025Proportionality, cembrano2022multidimensional, GaffkeDivisor, RoteMatrix}.

The apportionment problem is primarily motivated by the task of allocating seats in a legislature. However, ideas from randomized apportionment have found application in other domains, including peer selection~\cite{AzizPeerSelection} and affirmative action~\cite{AffirmativeAction}. One domain that presents particularly similar envy considerations to our setting is the selection of {citizens' assemblies}, which are randomly selected panels of ordinary citizens that convene to deliberate on an issue, typically to advise the legislature~\cite{CitizensAssemblies}. A very recent work by G\"olz et al.~\cite{CitySampling} considers the two-stage task of first selecting a set of cities proportional to their populations, then selecting the citizens within the city. They identify a similar notion of envy-freeness with respect to city size which they term \emph{ex post monotonicity} and give a sufficient condition under which it can be satisfied, which empirically holds in their specific application. Our paper is the first to systematically study the task of minimizing envy and provide worst-case bounds.

Further afield, randomized selection algorithms with correlation constraints have also been studied in fair division~\cite{IndivisibleGoodsSurvey}. This field is largely distinct from our setting in that agents have identical entitlements (i.e., standard quotas) but heterogeneous preferences over a complex set of goods being allocated (as opposed to common preferences for more seats). Similar techniques have been used to obtain fair randomized allocations, most notably the Birkhoff-von Neumann Theorem and its use in the classic \emph{probabilistic serial} algorithm~\cite{ProbabilisticSerial}.  Aziz et al.~\cite{BestOfBothWorlds} use this algorithm to obtain a \emph{best of both worlds} fairness guarantee, where an allocation is completely envy-free ex ante and envy-free up to the value of a single good in the envied player's allocation (EF1) ex post. This mimics the guarantees of the class of apportionment methods we study: ex ante proportionality and ex post quota.

\section{Preliminaries}\label{secPrelim}
We consider an apportionment problem with $n$ parties and a fixed house size $H$. Given vote counts $p_1, p_2, \dots, p_n$, each party $i$ is entitled to
$q_i = {p_iH}/{\sum_{j = 1}^n p_j}$
seats, which is called the \emph{standard quota} of party $i$. We may decompose these quotas as $q_i = f_i + r_i$, where $f_i = \lfloor q_i \rfloor$ is the party’s \emph{lower quota} and $r_i\in[0,1)$ is its \emph{residue}. After awarding each party its lower quota, exactly
$k = \sum_{i=1}^n r_i$
seats remain to be allocated. 

\subsection{Apportionment Rules and Obliviousness}\label{subsecRules}

An apportionment rule specifies a probability distribution over $k$-subsets, i.e., subsets $S\subseteq[n]$ of size $k$, where $i\in S$ indicates that party $i$ receives an additional seat. 
Equivalently, we may represent an allocation by a vector
$X\in\{0,1\}^n$ with $\sum_{i=1}^n X_i=k$, where $X_i=1$ whenever $i\in S$.
We restrict to rules satisfying the following two axioms:\\

\noindent {\bf Quota.} Each party $i\in [n]$ always receives $f_i = \lfloor q_i \rfloor$ or $\lceil q_i \rceil$ seats.\\

\noindent {\bf Ex-ante proportionality.} For each party $i\in [n]$, the probability of receiving an additional seat is exactly its residue, i.e., $\Pr(X_i = 1) = r_i$.\\

By construction, after awarding each party its lower quota, an apportionment rule satisfies the quota property. 
However, ex ante proportionality is not free, and the existence of such methods is a non-trivial mathematical result.
In what follows, we briefly recall two canonical apportionment rules studied in the literature and use them to motivate a general notion that will play a central role in our results.\\

\noindent\textbf{Grimmett’s systematic sampling.}
Grimmett~\cite{Grimmett} proposed an apportionment rule, based on the classic {\it systematic sampling} method from the statistics literature. 
After awarding each party $i$ its lower quota $f_i$, the method allocates the
remaining $k$ seats as follows. 
Arrange the residues $r_1,\dots,r_n$ as consecutive intervals on the real
line, with interval $i$ having length $r_i$. 
Then, draw $u$ uniformly at random from $[0,1)$, and award an additional seat to party $i$ if the shifted interval of party $i$ contains an integer. Since the interval $[u,u+k)$ contains exactly $k$ integers, this procedure always allocates exactly $k$ additional seats. Moreover, for each party $i$, the shifted interval contains an integer with probability exactly $r_i$, ensuring ex ante proportionality.\\

\noindent\textbf{Sampford rounding.}
This rule uses the sampling method introduced by Sampford~\cite{Sampford}. After awarding lower quotas, the rule selects a subset of $k$ parties using the following rejective procedure. First, sample a party $i_1$ with probabilities proportional to $r_i$. Then sample $k-1$ additional parties $i_2,\dots,i_k$ independently with replacement, where each party $i$ is chosen with probabilities proportional to $r_i/(1-r_i)$. If the $k$ sampled parties are distinct, select them; otherwise, repeat the procedure. This method induces a distribution over $k$-subsets with $\Pr(X_i=1)=r_i$ for all $i$. Sampford rounding has been studied extensively and is known to satisfy ex ante proportionality as well as negative association and additional correlation-based monotonicity properties in the randomized apportionment setting (e.g.,~\cite{MonotoneRandomizedApportionment}).

Both of the rules above share a common structural feature: their randomization depends only on the fractional parts of the quotas and ignores the deterministic seat allocations. 
This motivates the following notion.\\

\noindent{\bf Obliviousness.} An apportionment rule is \emph{oblivious} if its
randomization depends only on the vector of residues $(r_1,\dots,r_n)$, and
not on the lower quotas $(f_1,\dots,f_n)$.\\

An oblivious rule thus treats parties with identical residues symmetrically, regardless of their deterministic seat allocations. All randomized apportionment methods studied to date, including Grimmett’s method and Sampford rounding, are oblivious in this sense.
We remark that we use $r_i$ to denote both the fractional part of party $i$’s quota and its probability of receiving an additional seat.
For notational convenience, we index parties so that residues are in nondecreasing order:
\begin{equation}
	\label{eq:residue-order}
	0 \;\le\; r_1 \;\le\; r_2 \;\le\; \cdots \;\le\; r_n \;<\; 1,
\end{equation}

\subsection{Correlations and Envy}
The decomposition $q_i = f_i + r_i$ separates the deterministic and random components of the apportionment problem. Lower quotas $f_1,f_2,\ldots,f_n$ depend only on the vote counts, while the remaining $k$ seats are allocated through a randomized rounding step that selects exactly $k$ parties. 
Because quota and ex ante proportionality uniquely determine the marginal probabilities
$\Pr(X_i=1)=r_i$, all rules satisfying these axioms differ only in how they correlate the random variables $(X_1,\dots,X_n)$. 
We focus precisely on this correlation structure, since it is the only source of potential fairness violations. 
For any ordered pair of parties $(i,j)$ with $i\ne j$, we define
\begin{equation}
	\label{eq:pair-envy-global}
	e_{ij}= \Pr(X_i=0,\, X_j=1),
\end{equation}
the probability that party $j$ receives an additional seat while party $i$ does not. Intuitively, when parties $i$ and $j$ share the same lower quota and $r_i \ge r_j$, the quantity $e_{ij}$ measures the probability that party $i$ is {\it envious} of party $j$, in the sense that $i$ has a weakly higher entitlement yet receives fewer seats. 
We formalize envy by summing $e_{ij}$ over the appropriate pairs, as follows.

It will be useful to group together parties that share the same lower quota. Within such a group, all parties start from the same deterministic allocation and compete only for the additional seats assigned by the randomized rounding step. Specifically, we partition the set of parties into \(B\) \emph{equal-floor blocks} \(I_1, I_2, \dots, I_B \subseteq [n]\) pairwise disjoint, such that \(f_i=f_j\) whenever \(i,j\in I_b\). By convention, within each block \(I_b\) parties are ordered by nondecreasing residue order as in ~\eqref{eq:residue-order}. To aggregate envy, we account only for pairs of parties that belong to the same block, and only when a party with a higher residue fails to receive an additional seat while a lower-residue party in the same block does receive one---this is the only way that the seat allocation can vary non-monotonically with respect to the standard quotas.
The total expected envy is thus the following quantity:
\begin{equation}
	\label{eq:directional-sum}
	\ee\Big[\sum_{b=1}^B\sum_{i>j:i,j\in I_B}\ind{X_i=0,X_j=1}\Big]= \sum_{b=1}^B\sum_{i>j:i,j\in I_B} e_{ij}.
\end{equation}
When the total expected envy is zero, we say that the rule is {\it envy-free}.
In the single-floor case, all parties have the same lower quota, i.e, $B=1$, so any difference in the final allocation arises solely from the random assignment of the remaining $k=\sum_{i=1}^nr_i$ seats. 
In this case, the total expected envy corresponds to simply $\sum_{i>j} e_{ij}$.

A caveat with the expression in \eqref{eq:directional-sum} is that it breaks ties in an arbitrary way. It is a policy-maker's decision how to handle the case when two parties have the same standard quota: Perhaps neither party is justified in being envious if the other gets more seats, or perhaps both would be justified? We choose the formulation that the party numbered higher is entitled to weakly more seats because it is more natural from a mathematical perspective, enabling the proof of Theorem~\ref{thm:rule-indep}. 
Our impossibility result, Theorem~\ref{thm:oblivious-lower-bound}, gives instances in which standard quotas are distinct (as is highly likely in practice), and our algorithms easily extend to either alternative tie-breaking policy.

\subsection{Envy Aggregation Functions}\label{sec:envy-agg}
We now generalize Equation (\ref{eq:directional-sum}) to a set of alternative objective functions that may be more desirable depending on taste. Recall that an apportionment rule produces a random outcome $X \in \{0,1\}^n$ indicating which $k$ parties receive an additional seat. 
Let $\cS = \{S \subseteq [n] : |S| = k\}$ denote the set of feasible outcomes, where a subset $S$ corresponds to the event that $X_i=1$ if and only if $i \in S$. 
An apportionment rule can thus be identified with a probability distribution $p = (p(S))_{S \in \cS}$.
Under this representation, ex ante proportionality takes a
simple linear form: for each party $i \in [n]$, $
r_i = \Pr(X_i=1) = \sum_{S\in \cS: i\in S} p(S).$

Now suppose we have a partition $J_1,\ldots,J_L$ of $[n]$.
One natural partition is given by the equal-floor blocks, i.e., $L=B$ and $J_i=I_i$ for each $i$.
Another natural partition is just the trivial case of $L=1$ and $J_1=[n]$.
Apart from total envy, we can aggregate the envious pairs in the instance according to the partition $J$ to capture additional indicators for the apportionment rule.
In particular, the total envy due to partition $J$ corresponds to 
\begin{equation}
	\text{total expected envy}=\ee_{S\sim p}[\sigma_J(S)]=\textstyle\sum_{S\in \cS}\sigma_J(S)p(S),\label{eq:block-aggregation}
\end{equation}
where
$\sigma_J(S) = |\{(i,j) :\text{there exists }b\in [L]\text{ with } i>j,\ i,j \in J_b,\ i \notin S,\ j \in S\}|,$ i.e., the number of within-part inversions due to $S$, where a lower-residue party is selected while a higher-residue party in the same block is not.
Recall that parties are sorted according to non-decreasing residues.
The representation in \eqref{eq:block-aggregation} is particularly useful when considering other envy aggregation functions. 
For instance, the following two values are of interest:
\begin{align*}
	\text{Total envy variance}&=\text{Var}_{S\sim p}[\sigma_J(S)]=\ee_{S\sim p}\Big[\sigma_J^2(S)\Big]-\textstyle\Big(\sum_{S\in \cS}\sigma_J(S)p(S)\Big)^2,\\
	\text{Probability of {\bf not} being envy-free}&=\ee_{S\sim p}[\ind{\sigma_J(S)>0}]=\textstyle\sum_{S\in \cS:\sigma_J(S)>0}p(S).
\end{align*}
Observe that, for the total envy variance, taking the trivial partition $J_1=[n]$ recovers the variance of total envy in the instance, while taking the equal-floor blocks partition recovers the variance due to within-block inversions.
In Section~\ref{subsecMinEnvy}, we introduce an LP framework that allows us to minimize the total expected envy while remaining flexible enough to accommodate a more general class of envy-aggregation functions, including those above.

\section{Single-Floor Instances}\label{secSameFloor}
In this section, we study the apportionment problem in the special case where all parties have the same lower quota, i.e., $f_1=f_2=\cdots=f_n$; in particular, we have a single block $I_1=[n]$.
This setting serves as a warm-up that isolates the effect of random rounding without any interaction across different blocks. 
We show that in this case, the total expected envy is completely determined by the residue vector and is independent of the apportionment rule being used (Section~\ref{sec:single-floor-envy}). 
We further derive a tight upper bound on this quantity (Section~\ref{sec:single-floor-envy-tight}), which will later serve as a benchmark for the general multi-floor case.
\subsection{Total Expected Envy is Rule-Independent}\label{sec:single-floor-envy}
We begin with a lemma stating two identities that hold for any apportionment rule. They are used repeatedly in our analysis. 
\begin{lemma}\label{lem:repeated}
	For any apportionment rule, the following holds:
	\begin{enumerate}[label=\normalfont(\alph*)]
		\item \label{lem:unordered} 
		$\sum_{i<j}(e_{ij} + e_{ji})= k(n-k).$
		\item \label{lem:skew}For any \(i\ne j\), we have
		$e_{ij} - e_{ji}
		= \Pr(X_j=1) - \Pr(X_i=1) = r_j - r_i.$
	\end{enumerate}
\end{lemma}

\begin{proof}
	Recall that the outcome $X$ of a randomized rule satisfies $\sum_{i=1}^nX_i=k$.
	Observe that the number of unordered selected–unselected pairs is exactly \(k(n-k)\): there are \(k\) selected indices and \(n-k\) unselected indices.
	Therefore
	$\sum_{i<j}(\ind{X_i=1,X_j=0} + \ind{X_i=0,X_j=1}) = k(n-k).$
	Taking expectation yields \ref{lem:unordered}.
	For the second part, observe that
	\begin{align*}
		e_{ij} - e_{ji}&=\Pr(X_i=0,X_j=1)-\Pr(X_i=1,X_j=0)\\
		&= \Pr(X_j=1) - \Pr(X_i=1,X_j=1) - (\Pr(X_i=1) - \Pr(X_i=1,X_j=1))\\
		&= r_j - r_i,
	\end{align*}
	where the last equality holds since the rule is ex ante proportional.
\end{proof}

We remark that~\ref{lem:skew} shows that envy is biased toward lower residues in direct proportion to the gap in marginals.
To combine the two identities it is convenient to write each pair $(e_{ij}, e_{ji})$ in terms of its symmetric and antisymmetric parts: for $i\neq j$, let $S_{ij} = e_{ij} + e_{ji}$ and $\Delta_{ij} = e_{ij} - e_{ji}$. By Lemma~\ref{lem:repeated}\ref{lem:unordered} and~\ref{lem:repeated}\ref{lem:skew}, we have
$\sum_{i<j} S_{ij} = k(n-k),$ and
$\Delta_{ij} = r_j-r_i.$
These identities allow us to express the total expected envy in closed form as a function of the residue vector alone.
Furthermore, in the following theorem, we show that the total expected envy depends only on the residues $r_1,\ldots,r_n$, not on the randomized rule.
\begin{theorem}
	\label{thm:rule-indep}
	Consider a single-floor instance with residues $0\le r_1\le\cdots\le r_n<1$. Then, for any apportionment rule, the total expected envy is equal to
	\begin{equation}
		\label{eq:E-closed-form}
		\frac{1}{2}k(n-k) + \frac{1}{2}\sum_{t=1}^n r_t(n-2t+1).
	\end{equation}
\end{theorem}
\begin{proof}
	Using the symmetric/antisymmetric decomposition,
	\[
	e_{ji} \;=\; \frac{S_{ij} + \Delta_{ji}}{2}
	\;=\; \frac{S_{ij} + (e_{ji}-e_{ij})}{2}
	\;=\; \frac{S_{ij} + r_i-r_j}{2},
	\]
	where the last equality uses Lemma~\ref{lem:repeated}\ref{lem:skew}. 
	Then, 
	\[
	\sum_{i>j} e_{ij}= \sum_{i<j} e_{ji}
	= \frac{1}{2}\sum_{i<j} S_{ij} + \frac{1}{2}\sum_{i<j}(r_i-r_j)=\frac{1}{2}k(n-k) + \frac{1}{2}\sum_{i<j}(r_i-r_j),
	\]
	where the last equality follows from Lemma~\ref{lem:repeated}\ref{lem:unordered}. 
	For the second sum, $r_t$ appears positively whenever $i=t$ and $j>t$ (there are $n-t$ such pairs), and negatively whenever $j=t$ and $i<t$ (there are $t-1$ such pairs). 
	Hence
	$\sum_{i<j}(r_i-r_j) = \sum_{t=1}^n r_t((n-t)-(t-1))
	= \sum_{t=1}^n r_t(n-2t+1).$
\end{proof}
\subsection{Tightness and Party Envy}\label{sec:single-floor-envy-tight}
As an immediate consequence of Theorem~\ref{thm:rule-indep}, we obtain a sharp upper bound on the total expected envy in the single-floor case. The bound characterizes the worst possible instance and identifies when it is attained.
\begin{proposition}\label{cor:half-bound}
	In the space of single-floor instances with residues $0\le r_1\le\cdots\le r_n<1$, any apportionment rule has total expected envy at most $k(n-k)/2$, and this upper bound is tight for each $n$ and $k<n$.
	In particular, the total expected envy is at most $n^2/8$.
\end{proposition}
\begin{proof}
	We start by observing that the second summation in \eqref{eq:E-closed-form} is maximized whenever $\sum_{t=1}^nr_t\cdot t$ is minimized.
	Since the sequence $(r_t)_{t=1}^n$ is nondecreasing, Chebyshev's sum inequality implies that $\sum_{t=1}^nr_t\cdot t\ge (\sum_{t=1}^nr_t)(\sum_{t=1}^nt)/n=k(n+1)/2$.
	Therefore, 
	$$\sum_{t=1}^n r_t(n-2t+1)\le (n+1)\sum_{t=1}^nr_t-2\sum_{t=1}^nr_t\cdot t=(n+1)k-2k(n+1)/2=0,$$
	which implies that the total expected envy is at most $k(n-k)$.
	For each $n$ and $k<n$, this inequality is tight by considering an instance where the residues are $r_i=k/n$ for each $i\in [n]$.
	Finally, the upper bound of $n^2/8$ is direct from the fact that $x\mapsto x(n-x)/2$ is maximized when $x=n/2$.
	This value is also attained when $r_i=1/2$ for each $i\in [n]$.
\end{proof}

Together with Theorem~\ref{thm:rule-indep}, this bound shows that in the single-floor case, the residues alone determine both the exact value and the maximum possible magnitude of total expected envy. 
In the next section, we move to multi-floor blocks, where this clean rule independence no longer holds and envy minimization becomes a genuine optimization problem. 

Beyond the total expected envy, one can also derive closed-form expressions for the envy of individual parties, including the highest-residue party and the aggregate envy of the middle parties. 
For this, let $e_1=0$ and let $e_i=\sum_{j<i}e_{ij}$, for each $i\in \{2,\ldots,n\}$, be the expected envy of party $i$.
In particular, observe that the total expected envy equals $\sum_{i=1}^ne_i$.
We have the following proposition.
\begin{proposition}\label{prop:envy-parties}
	In the space of single-floor instances with residues $0\le r_1\le\cdots\le r_n<1$, for any apportionment rule we have $e_n= k(1-r_n).$
\end{proposition}
\begin{proof}
	From the definition of $e_n$, we have
	\[
	e_n 
	= \sum_{j<n} e_{nj}
	= \sum_{j<n}(r_j - \Pr(X_n=1,X_j=1))
	= k-r_n - \sum_{j<n} \Pr(X_n=1,X_j=1),
	\]
	where the second equality follows from $r_j=\Pr(X_j=1)$, and the last one from
	$\sum_{j=1}^n r_j = k$.
	On the other hand, we have
	$\sum_{j<n} X_iX_j
	= X_n\sum_{j< n} X_j
	= X_n(k - X_n)
	= (k-1)X_n,$
	since $\sum_{j=1}^nX_j=k$ and $X_n^2=X_n$. 
	Taking expectations gives
	$\sum_{j< n}\Pr(X_n=1,X_j=1)=\ee[(k-1)X_n]=(k-1)r_n.$
\end{proof}

\section{Multi-Floor Instances}\label{secMultifloor}

We now consider the general case of instances with multiple equal-floor blocks. 
In contrast to the single-floor setting, interactions across different lower quotas break the rule-independence of total expected envy: for a fixed residue vector, different apportionment rules can lead to substantially different levels of envy; e.g., the example in Section~\ref{subExample} about Sampford's and Grimmett's methods.

In Section~\ref{subsec:oblivious-multifloor}, we begin by studying the limitations of oblivious apportionment rules, those whose randomization depends only on the residues and ignores lower quotas. 
We show that in the multi-floor setting, such rules can be arbitrarily suboptimal, incurring quadratic envy even when envy-free solutions exist.

We then turn to the design of optimal (non-oblivious) apportionment rules. 
In Section~\ref{subsecLP1}, we show that it is possible to decide whether a given instance admits an envy-free apportionment rule by solving a network flow. 
In Section~\ref{subsecMinEnvy}, we handle the case when envy-freeness does not necessarily hold by formulating an exact linear program that minimizes total expected envy. 
Although this program has exponentially many variables, we show that it can be solved in polynomial time using a carefully designed separation oracle.

\subsection{Oblivious Apportionment Rules}\label{subsec:oblivious-multifloor}

We show that oblivious apportionment rules can perform extremely poorly in the multi-floor setting. 
Recall that a rule is oblivious if its randomization depends only on the residue vector and ignores the lower quotas. While such rules are natural and encompass all previously studied randomized methods, we show that they can incur quadratic envy even in instances where envy-free randomized rules exist. Combined with the universal quadratic upper bound in Proposition~\ref{cor:half-bound}, this shows that oblivious rules are as bad as possible up to constant factors.
A notable feature of our construction is that the adversary is non-adaptive, namely, we construct a distribution over adversarial instances that fools any randomized algorithm that outputs an apportionment rule. 
This implies, in particular, that any fixed rule incurs quadratic envy on some instance.

\begin{theorem}
	\label{thm:oblivious-lower-bound}
	For arbitrarily large $n$, there is a distribution over apportionment instances with $n$ parties with distinct standard quotas for which the following holds:
	\begin{enumerate}[label=\normalfont(\roman*)]
		\item\label{itmLowerBoundConstructionEF} An envy-free apportionment exists with probability one.
		\item\label{itmLowerBoundConstruction1OB} Any oblivious apportionment rule incurs expected envy $\Omega(n^2)$.
	\end{enumerate}
\end{theorem}

\begin{proof}
	Our adversarial distribution is defined as follows, for $n$ divisible by 4. Let $\varepsilon = {1}/{(10n)}$. 
	Pick a subset $T \subseteq [n/2]$ of size ${n}/{4}$ uniformly at random. We denote by $\overline{T}$ the complement of $T$ within $[n/2]$, i.e., $\overline{T} := [n/2] \setminus T$. 
	The apportionment instance is defined by the following standard quotas:
	$$q_i = \begin{cases}\frac32 + (i - 1 - \frac{n}{2})\varepsilon & \text{if } i \in T,\\ \frac12 + (i - 1 - \frac{n}{2})\varepsilon & \text{if } i \in [{n}/{2}] \setminus T, \\ 1 + (i - \frac{n}{2})(1 + \varepsilon) &\text{if } i > {n}/{2}.\end{cases}$$
	
	Figure~\ref{figLowerBoundExample} shows an example where $n = 8$.
	Observe that, since $n\varepsilon < 1$, all standard quotas are distinct. 
	Also, the sum of residues is
	$$k = \sum_{i = 1}^{n / 2} \left(\frac12 + \left(i - 1 - \frac{n}{2}\right)\varepsilon\right) + \sum_{i = \frac{n}{2} + 1}^{n} \left(i - \frac{n}{2}\right)\varepsilon = \frac{n}{4},$$
	where the first sum collects the residues in $T$ and $[n/2]\setminus T$, and the second sum collects the residues for $i>n/2$; observe that the $\varepsilon$ part of the $j\tth$ term of the first sum cancels out with the $j\tth$-last term of the second sum, leaving only ${n}/{2}$ copies of $1/2$.
	
	\ipncm{.4}{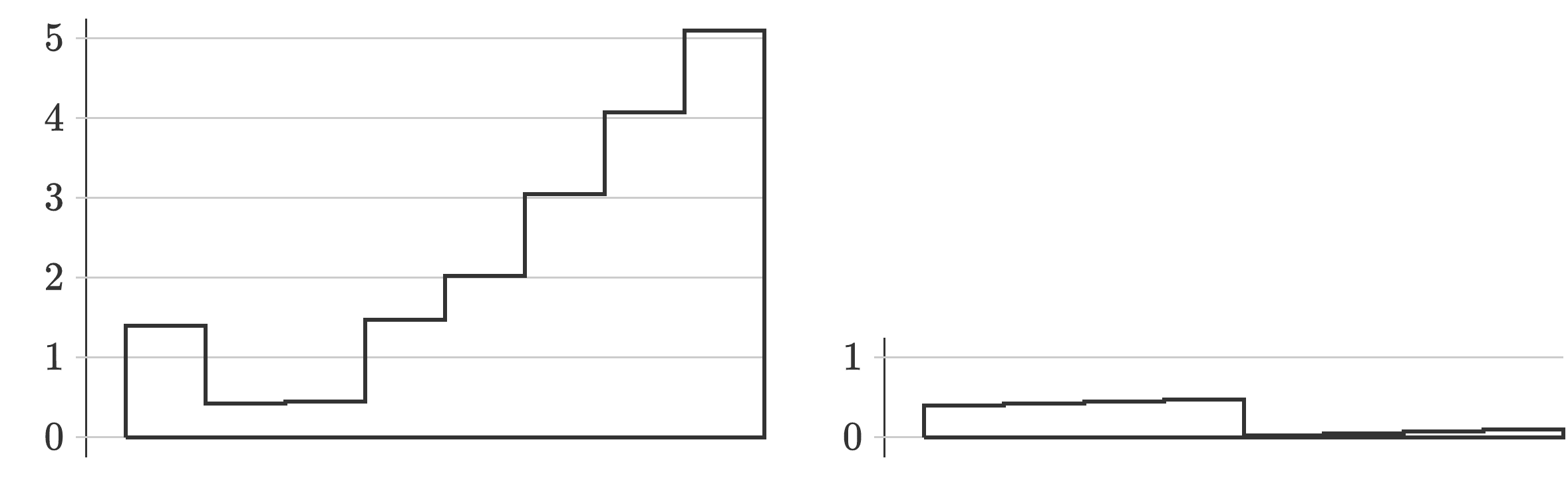}{\label{figLowerBoundExample} Construction with $n = 8$, $k = 2$, and $T = \{1, 4\}$. On the left is shown the standard quotas of each party, and on the right is shown the residues, which cluster around $1/2$ and zero. The tiny steps have height $\varepsilon$. There is an envy-free apportionment that puts roughly half probability on $\{1, 4\}$ and half probability on $\{2, 3\}$, but an oblivious rule only sees the residues on the right, so it cannot identify this partition.}
	
	First let us show that an envy-free randomized apportionment exists. We claim that the following distribution over sets $S$ of size $k = {n}/{4}$ is envy-free and proportional; see Figure~\ref{figStaircase} for an illustration of case \ref{itmStaircase}:
	
	\begin{enumerate}[label=(\arabic*)]
		\item\label{itmT} With probability ${(1 - n\varepsilon)}/{2}$, select $S = T$.
		\item\label{itmTBar} With probability ${(1 - n\varepsilon)}/{2}$, select $S = \overline{T}$.
		\item\label{itmStaircase} With remaining probability $n \varepsilon$, pick a uniformly random index $j \sim [n/4]$, then select one of the four sets, each with probability $1/4$:
		\begin{enumerate}[label=\normalfont(\Alph*)]
			\item\label{itmStaircaseA} $S = \{\frac{n}{4} + 1 + j, \dots, \frac{n}{2}\} \cup \{n + 1 - j, \dots, n\}$.
			\item\label{itmStaircaseB} $S = \{\frac{3n}{4} + 1 - j, \dots, \frac{3n}{4}\} \cup \{\frac{3n}{4} + 1 + j, \dots, n\}$.
			\item\label{itmStaircaseC} $S = (\{\frac{n}{4} + 2 - j, \dots, \frac{n}{2}\} \cap T) \cup \{\frac{3n}{4} + 1, \dots, \frac{3n}{4} + \abs{\{1, \dots, \frac{n}{4} + 1 - j\} \cap T}\}$.
			\item\label{itmStaircaseD} $S = (\{\frac{n}{4} + 2 - j, \dots, \frac{n}{2}\} \cap \overline{T}) \cup \{\frac{3n}{4} + \abs{\{1, \dots, \frac{n}{4} + 1 - j\} \cap \overline{T}}, \dots, n + 1 - j\}$.
		\end{enumerate}
	\end{enumerate}
	
	\ipncm{.4}{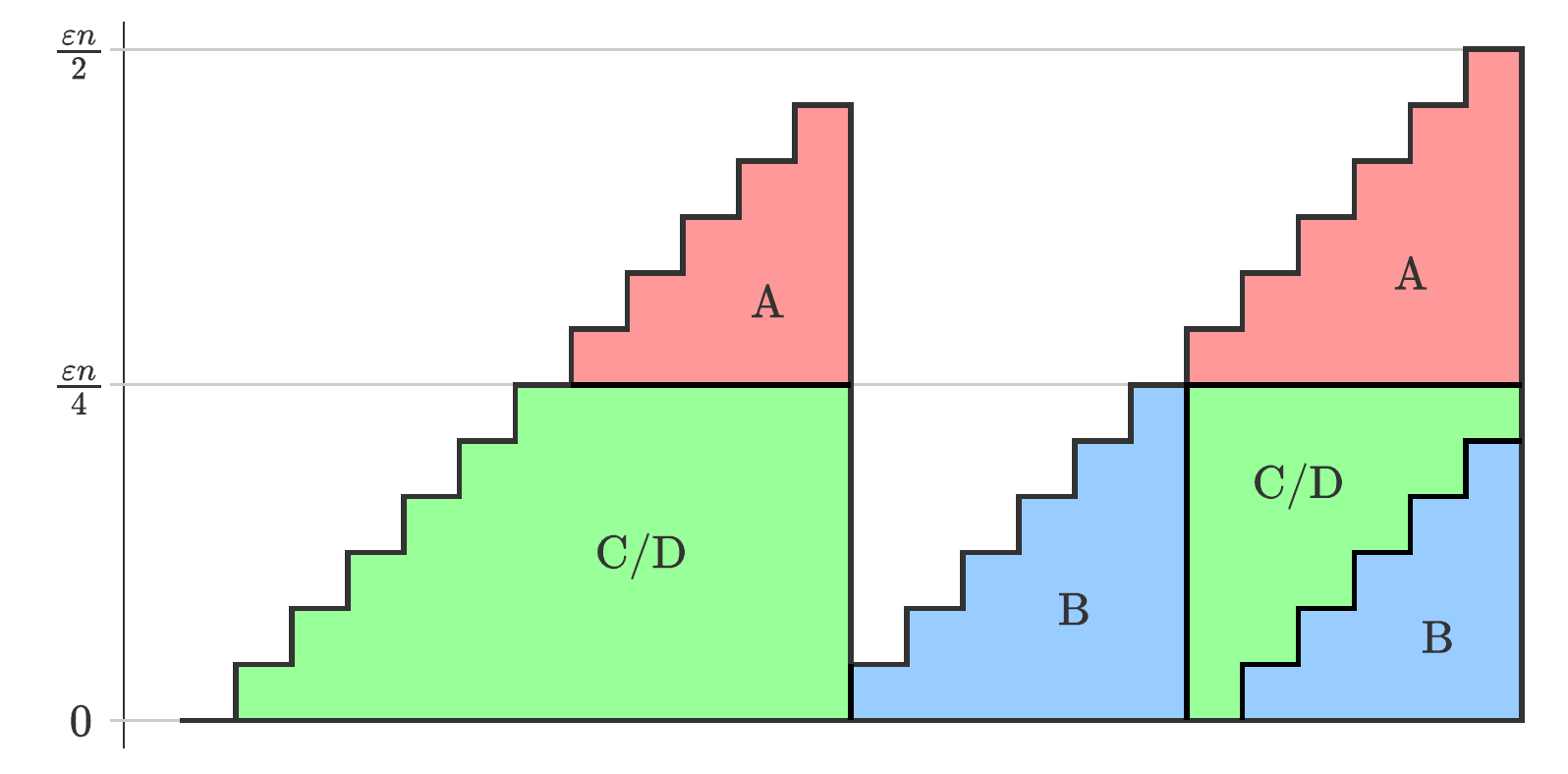}{\label{figStaircase}Illustration of Case \ref{itmStaircase} of the envy-free randomized apportionment. In this example, $n = 12$, and shown are the remaining residues of each party after taking away probability mass according to Cases \ref{itmT} and \ref{itmTBar}. The remaining probability mass is partitioned according to the sub-cases of case \ref{itmStaircase} in which they are bundled into additional sets of size $k = {n}/{4}$.}
	
	The sets selected in Cases \ref{itmT} and \ref{itmTBar} are clearly envy-free, because they consist of all parties with lower quotas 1 and 0, respectively. After subtracting probability ${(1 - n \varepsilon)}/{2}$ from each of the first ${n}/{2}$ residues, the remaining residues of each party are plotted in Figure~\ref{figStaircase}. Case 3 essentially pairs up the corresponding regions, always selecting $k = {n}/{4}$ parties at a time so that all $\varepsilon \times \varepsilon$ blocks are accounted for. For instance, case \ref{itmStaircaseA} takes the $j$ blocks off the $j\tth$-top row of the A-region on the right and combines them with the ${n}/{4} - j$ blocks from the $j\tth$ bottom row of the A-region on the left. The B-regions are analogously paired in Case \ref{itmStaircaseB}. Cases \ref{itmStaircaseC} and \ref{itmStaircaseD} collectively pair the $j\tth$-top row in both regions, which has ${n}/{2}$ blocks. Notice that each such set $S$ consists of the top residues in $T$ and $\overline{T}$, so there is never any envy.
	
	Now fix an oblivious apportionment rule. For the remainder of this proof, all probabilities are taken over both the random set $T$ chosen by the adversary and the random set $S$ selected by the rule. Note that $T$ and $S$ are independent random variables. As usual, we denote by $X_i$ the indicator of the event that $i \in S$. 
	We show that the expected envy is at least $\frac{n^2}{259200}$.
	First, we observe that
	\begin{equation}\label{equSContainsSmallValues}
		\Pr\left[\abs{S \cap \left[\frac{n}{4}\right]} \geq \frac{n}{20}\right] \geq \frac{1}{4},
	\end{equation}
	for otherwise we would have
	\begin{align*}
			\frac{9n}{80} &= \frac{n}{4} \cdot \frac{1 - n \varepsilon}{2} \snc{n\varepsilon = \frac{1}{10}}\\
			&< \sum_{i = 1}^{n/4} \ee\left[X_i\right]\\
			&= \ee\left[\sum_{i = 1}^{n/4} X_i\right]\\
			&= \ee\left[\abs{S \cap \left[\frac{n}{4}\right]}\right]\\
			&\leq \Pr\left[\abs{S \cap \left[\frac{n}{4}\right]} \geq \frac{n}{20}\right] \cdot \frac{n}{4} + \Pr\left[\abs{S \cap \left[\frac{n}{4}\right]} < \frac{n}{20}\right] \cdot \frac{n}{20}\\
			&\leq \frac14 \cdot \frac{n}{4} + 1 \cdot \frac{n}{20} \stext{assuming Equation (\ref{equSContainsSmallValues}) is false}\\
			&= \frac{9n}{80},
	\end{align*}
	which is a contradiction.
	
	Now fix any $S$ containing at least ${n}/{20}$ parties from $[{n}/{4}]$; call these parties $A \subseteq [{n}/{4}]$. Since $\abs{S} = {n}/{4}$, this means that there are also at least ${n}/{20}$ parties in $\{{n}/{4} + 1, \dots, {n}/{2}\}$ that $S$ does \emph{not} contain. Consider the random set $T$, which is still a uniformly random subset of $[{n}/{2}]$, even after conditioning on $S$, by independence. 
	With probability at least $1/2$, at least $1/3$ of the elements of $T$ are from $[{n}/{4}]$ and at least $1/3$ are from $\{{n}/{4} + 1, \dots, {n}/{2}\}$.\footnote{The $\frac13$ could be replaced by any constant strictly less than 1/2 and both events actually occur with high probability. As we are not concerned with constant factors, we only require a constant probability lower bound of $\frac12$.} Conditioning on any such realization of $\abs{T \cap [{n}/{4}]} \in [\frac13\cdot \frac{n}{4}, \frac23 \cdot \frac{n}{4}]$,
	\begin{align*}
		\Pr\left[\abs{T \cap A} \geq \frac{n}{60} \txt{ and } \abs{T \cap B} \geq \frac{n}{60}\right] &= \Pr\left[\abs{T \cap A} \geq \frac{n}{60}\right]\cdot\Pr\left[\abs{T \cap B} \geq \frac{n}{60}\right]
		\geq \frac12 \cdot \frac12 = \frac14.
	\end{align*}
	Thus, with probability at least $\frac14 \cdot \frac12 \cdot \frac13 \cdot \frac13 = \frac{1}{72}$, $S$ contains ${n}/{60}$ parties in $T \cap A$ but does not contain at least ${n}/{60}$ parties in $T \cap B$, in which case the latter parties are envious. Thus, the total expected envy is at least $\frac{1}{72} \cdot \frac{n}{60} \cdot \frac{n}{60} = \frac{n^2}{259200}$.
\end{proof}

\subsection{Envy-Freeness Detection via Network Flows}\label{subsecLP1}
In this section, we address the envy-freeness question in the multi-floor setting: when does there exist an apportionment rule with zero total expected envy? Recall that within a block, zero envy can only be achieved if the selected parties are precisely those with the highest residues.

Formally, consider the blocks $I_1,\dots,I_B$, where block $b$ has size $m_b = |I_b|$, and parties within each block are indexed according to nondecreasing residues. A necessary and sufficient condition for zero envy is that, in every outcome and every block $b$, the selected parties in $I_b$ are exactly the top $\ell$ parties for some $\ell \in \{0,\dots,m_b\}$. 
This observation reduces the space of feasible outcomes drastically since, instead of choosing arbitrary subsets of size $k$, a rule with zero total expected envy only needs to decide, for each block, \emph{how many} parties to select. 
The remaining difficulty is to coordinate these per-block choices so that exactly $k$ parties are selected overall, and ex ante proportionality is met. 
We model this problem as a network-flow linear program.

\paragraph{\bf Network flow construction.} We introduce variables
$f_{b,s,\ell} \ge 0,$
which represent the probability mass of outcomes in which, after processing blocks $1,\dots,b-1$, a total of $s$ parties have been selected, and block $b$ contributes exactly $\ell$ additional parties (namely, the top $\ell$ in residue order). Since the final outcome must select exactly $k$ parties, feasibility requires $s+\ell \le k$. 
These variables define a flow that moves from block to block, keeping track of the cumulative number of selected parties. 
At each block, the flow branches according to the number of parties selected in that block, and the proportionality constraints restrict how that number is distributed across blocks.
The feasibility program is given by the following constraints:

\begin{align}
	\sum_{\ell=0}^{\min(m_1,k)} f_{1,0,\ell} &= 1,\tag*{\normalfont{\mbox{[EF-LP]}}}\label{form:EF}\\
	\sum_{\ell=0}^{\min(m_1,k-s)} f_{1,s,\ell} &= 0\quad \text{ for every }s\in \{1,\ldots,k\},\label{LP:init2}\\
	\sum_{\ell=0}^{\min(m_b,k-s)} f_{b,s,\ell}-
	\hspace{-.4cm}\sum_{\ell'=0}^{\min(m_{b-1},s)} f_{b-1,s-\ell',\ell'}&=0\quad \text{ for every }s\in \{0,1,\ldots,k\}\text{ and }b\in \{2,\ldots,B\},\label{LP:conservation}\\
	\sum_{\ell'=0}^{\min(m_B,k)} f_{B,k-\ell',\ell'} &= 1,\label{LP:termination}\\
	\sum_{s=0}^k \sum_{\ell = m_b - t + 1}^{\min(m_b,k-s)} f_{b,s,\ell}
	&= r_t\quad\text{ for every }b\in\{1,\ldots,B\}\text{ and }t\in \{1,\ldots,m_b\},\label{LP:exanteprop}\\
	f&\ge 0.\notag
\end{align}
Constraints \eqref{LP:init2} initialize the probability flow. 
They force all probability mass to start before the first block with zero selected
parties. 
Thus, before processing block $b=1$, the only reachable state is $s=0$.
Constraint \eqref{LP:conservation} is the flow-conservation constraint: for every later block, the mass leaving a cumulative state equals the
mass entering that state from the previous block. 
Constraint \eqref{LP:termination} forces all probability mass to end in outcomes
that select exactly \(k\) parties in total. 
Hence, every path carrying positive
flow corresponds to a feasible apportionment outcome.
Finally, \eqref{LP:exanteprop} enforces ex ante proportionality; observe that for a block $b$ and a position $t \in \{1,\dots,m_b\}$, selecting the top $\ell$ parties in block $b$ includes party $t$ if and only if
$\ell \ge m_b - t + 1$. 
We have the following proposition.

\begin{proposition}\label{prop:envyfreenes}
	Given an apportionment instance, there exists an envy-free apportionment rule if and only if the linear program \ref{form:EF} is feasible.
\end{proposition}

\begin{proof}
	If the linear program is feasible, since the network is acyclic, any feasible flow can be decomposed into a distribution over paths; see, e.g., \cite[section 3.5]{magnanti1993network}.
	Each path specifies how many parties are selected from each block, and the corresponding outcome selects the top
	parties within every block. Therefore, the induced rule has zero total expected
	envy and satisfies ex ante proportionality.
	Conversely, any envy-free apportionment rule induces feasible flow values by
	setting
	$f_{b,s,\ell} = \Pr(S_{b-1}=s,\; K_b=\ell),$
	where $S_{b-1}$ is the number of parties selected before block $b$ and $K_b$ the number selected in $b$.
\end{proof}

\begin{figure}[H]
	\centering
	\includegraphics[scale=.62]{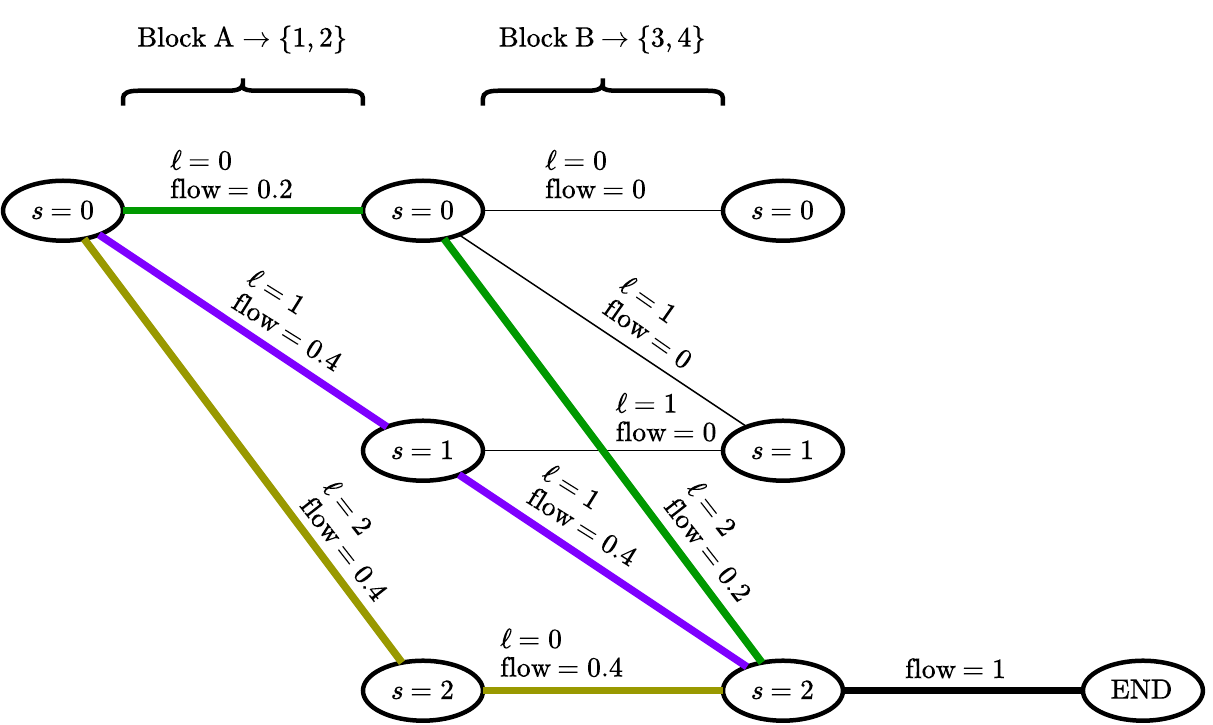}
	\caption{Flow network for the introductory example. Nodes are labeled by the cumulative number of additional seats~$s$. Colored edges indicate the three paths carrying nonzero flow: the brown path selects both parties from Block~$A$ (allocation $(2,2,3,3)$), the purple path selects one from each block (allocation $(1,2,3,4)$), and the green path selects both from Block~$B$ (allocation $(1,1,4,4)$). Gray edges are structurally valid but carry zero flow.}
	\label{fig:flow-example}
\end{figure}

The number of variables and constraints in \ref{form:EF} is $O(nk)$.
Therefore, it can be solved efficiently using standard linear programming methods.
We illustrate the flow formulation on the example from Section~\ref{subExample}, reproduced in Table~\ref{tabIntroExample} for convenience. 
There are $n = 4$ parties, $H = 10$ seats, and the residues sum to $k = 0.4 + 0.8 + 0.2 + 0.6 = 2$, so two additional seats must be distributed. The parties partition into two equal-floor blocks:
\begin{itemize}
	\item Block $A = \{1, 2\}$ with lower quota $1$, residues $(0.4, 0.8)$;
	\item Block $B = \{3, 4\}$ with lower quota $3$, residues $(0.2, 0.6)$.
\end{itemize}
Within each block, parties are sorted by nondecreasing residue.
The network has three layers of nodes, corresponding to the state before processing any block, the state after Block~$A$, and the state after Block~$B$. Each node is labeled by the cumulative number of additional seats $s$ assigned so far. Since $k = 2$, the possible values are $s \in \{0, 1, 2\}$. An edge from node $(s)$ in column $b$ to node $(s + \ell)$ in column $b+1$ represents the decision to select the top-$\ell$ parties from block $b$, and carries flow equal to the probability mass assigned to that choice. The network is depicted in Figure~\ref{fig:flow-example}.

Recall that an outcome has zero envy if and only if, within each block, the selected parties are exactly the top-$\ell$ in residue order for some $\ell$. In Block~$A$, the top-$1$ consists of party~$2$ alone (the higher residue), while the top-$2$ includes both parties~$1$ and~$2$. In Block~$B$, the top-$1$ is party~$4$ and the top-$2$ includes both parties~$3$ and~$4$. Since each block has size~$2$ and the total selection is $k = 2$, there are exactly three zero-envy outcomes, corresponding to the three ways of splitting two selections across two blocks:
\begin{center}
	\begin{tabular}{cccl}
		\toprule
		$\ell_A$ & $\ell_B$ & Selected parties & Seat allocation \\
		\midrule
		2 & 0 & $\{1, 2\}$ & $(2, 2, 3, 3)$ \\
		1 & 1 & $\{2, 4\}$ & $(1, 2, 3, 4)$ \\
		0 & 2 & $\{3, 4\}$ & $(1, 1, 4, 4)$ \\
		\bottomrule
	\end{tabular}
\end{center}
Each outcome corresponds to a source-to-sink path in the network: the first passes through $s = 0 \to 2 \to 2$, the second through $s = 0 \to 1 \to 2$, and the third through $s = 0 \to 0 \to 2$.
Then, the LP is feasible and the following defines an envy-free apportionment rule:
\[
\text{Seats allocation} =
\begin{cases}
	(2, 2, 3, 3) & \text{with probability } 0.4,\\
	(1, 2, 3, 4) & \text{with probability } 0.4,\\
	(1, 1, 4, 4) & \text{with probability } 0.2.
\end{cases}
\]


\subsection{Envy Aggregation via Linear Programming Duality}\label{subsecMinEnvy}
When an envy-free proportional rule does not exist, a natural objective is to minimize the total expected envy.
In this section, we consider a general linear programming framework to handle this objective and, more broadly, we introduce the class of {\it block-aggregable} functions, which are amenable to minimizing their expectation using the representation in Section~\ref{sec:envy-agg}.

Recall that apportionment rules are given by distributions over the subsets of parties that receive an additional seat, i.e., $S\mapsto p(S)$ is the probability of selecting $S\in \cS=\{T\subseteq[n]: |T|=k\}$.
Given a function $\mu:\cS\to \mathbb R$, consider the following linear program:
\begin{align}
	\text{minimize }  \quad\textstyle\sum_{S\in\cS} \mu(S)&p(S) \tag*{\normalfont{\mbox{[MinLP]}}}\label{form:minLP}\\
	\text{subject to }\quad 
	\textstyle\sum_{S\in \cS:i\in S} p(S) &= r_i \quad \text{ for every } i\in[n],\notag\\
	p(S) &\ge 0\quad \text{ for every }S\in \cS.\notag
\end{align}
We remark that every feasible solution $p$ in \ref{form:minLP} satisfies $\sum_{S\in \cS}p(S)=1$. Indeed, for such $p$, we have
$k=\sum_{i=1}^n r_i = \sum_{i=1}^n \sum_{S\in \cS:i\in S} p(S) = \sum_{S\in\cS} |S|p(S) = k \sum_{S\in\cS} p(S).$

The linear program is conceptually simple but has exponentially many variables, one for each feasible outcome $S \in \cS$. 
A standard approach in these cases is to consider the dual problem, study the complexity of the separation problem, and then solve efficiently using the ellipsoid method; see, e.g.,~\cite{grotschel2012geometric}.
In this case, the dual of \ref{form:minLP} is the following program:
\begin{align}
	\text{maximize }  \quad\textstyle\sum_{i=1}^n r_i&y_i \tag*{\normalfont{\mbox{[DualLP]}}}\label{form:dualLP}\\
	\text{subject to }\quad 
	\textstyle\sum_{i\in S}y_i &\le \mu(S) \quad \text{ for every } S\in \cS.\notag
\end{align}
Then, the separation problem is the following:
\begin{equation}
	\max_{S\in\cS}\Big\{\textstyle\sum_{i\in S} y_i - \mu(S) \Big\}.\tag*{\normalfont{\mbox{[SEP]}}}\label{eq:oracle-max}
\end{equation}
Our goal is to solve the problem \ref{eq:oracle-max} efficiently, in particular, using dynamic programming. 
Recall that given a partition $J=\{J_1,\cdots, J_L\}$, we introduce in Section~\ref{sec:envy-agg} the function
$$\sigma_J(S)=|\{(i,j) :\text{there exists }b\in [L]\text{ with } i>j,\ i,j \in J_b,\ i \notin S,\ j \in S\}|\text{ for every }S\subseteq [n],$$ i.e., the number of within-part inversions due to $S$, where a lower-residue party is selected while a higher-residue party is not.
In particular, we have 
\begin{equation}
	\textstyle \sigma_J(S)=\sum_{b=1}^L\sigma_J(S\cap J_b)\quad \text{for every }S\subseteq [n].\label{eq:decomposability}
\end{equation}
We say that $\mu:\cS\to \mathbb R$ is {\it block-aggregable} respect to $J$ if there exists $\psi:\mathbb N\to\mathbb R$ such that $$\mu(S)= \textstyle\psi(\sigma_J(S))\text{ for every }S\in \cS.$$
We suppose in what follows that $\psi:\{0,1,\dots,n^2\}\to\mathbb R$ can be evaluated in polynomial time.
The envy aggregations considered in Section~\ref{sec:envy-agg} are captured by block-aggregable functions.
For the case of total expected envy, it is sufficient to take $\psi(x)=x$, for the probability of not being envy-free, we can take $\psi(x)=\ind{x>0}$, and for the variance of total expected envy, we can take $\psi(x)=x^2$ for handling the first term.
In the following result, we show that \ref{form:minLP} can be solved efficiently for block-aggregable functions, using dynamic programming for the separation problem.
\begin{theorem}\label{thm:DPblock}
	Let $I$ be a partition of $[n]$, and let
	$\mu:\cS\to \mathbb R$ be block-aggregable with respect to
	$I$. 
	Then, \ref{form:minLP} can be solved in polynomial time.
\end{theorem}

\begin{proof}
	It is sufficient to show that the separation problem \ref{eq:oracle-max} can be solved in polynomial time.
	We show that for block-aggregable functions, this can be done using dynamic programming thanks to the decomposability property \eqref{eq:decomposability}.
	For each part $J_b$, let $i_1<i_2<\cdots<i_{m_b}$ be its elements, with $m_b=|J_b|$, and let $M_b=\binom{m_b}{2}$. 
	We first compute an auxiliary table for this set $J_b$. For $q\in \{0,\dots,\min\{k,m_b\}\}$ and $u\in \{0,\dots,M_b\}$,
	define
	\[F_b(q,u)=\max\Big\{\textstyle\sum_{i\in T} y_i:T\subseteq J_b, |T|=q, \sigma_J(T)=u\Big\},\]
	where the maximum over an empty feasible set is interpreted as $-\infty$.
	We compute $F_b$ by dynamic programming. Let $H[r,q,u]$ be the maximum
	value of $\sum_{i\in T} y_i$
	over sets $T\subseteq\{i_1,\dots,i_r\}$ such that $|T|=q$ and
	$|\{(i_s,i_t): s>t,\ i_s\notin T,\ i_t\in T\}|=u$.
	Initialize $H(0,0,0)=0$, and $-\infty$ otherwise.
	For $r=1,\dots,m_b$, the recurrence is
	\[
	H(r,q,u)=\max\Big\{H(r-1,q-1,u)+y_{i_r},H(r-1,q,u-q)\Big\}.
	\]
	To justify the recurrence, consider two cases. If
	$i_r\in T$, then before processing $i_r$ we must have selected $q-1$
	elements from $\{i_1,\dots,i_{r-1}\}$. Moreover, selecting $i_r$ does not
	create any new envious pair, because the larger element $i_r$ is selected. 
	Hence, the number of pairs remains $u$, and the contribution to the objective increases by $y_{i_r}$.
	
	If instead $i_r\notin T$, then the same $q$ selected elements already
	belonged to $\{i_1,\dots,i_{r-1}\}$. Since $i_r$ is larger than all
	previous elements in the block, it creates exactly one new pair with each
	previously selected element. Therefore, exactly $q$ new pairs are created, and in order to end with $u$ pairs after processing $i_r$, we must have had $u-q$ pairs before processing $i_r$. 
	At the end of the dynamic program, we set $F_b(q,u)=H(m_b,q,u).$
	This computation can be done in polynomial time.
	
	We now define the global dynamic program. 
	Let $M=\sum_{b=1}^B M_b$.
	For the values $b\in \{0,\dots,B\}$, $t\in \{0,\dots,k\}$, and $a\in \{0,\dots,M\}$,
	let 
	$$D(b,t,a)=\max\Big\{\textstyle \sum_{\ell=1}^b \sum_{i\in S_\ell} y_i:S_\ell\subseteq J_\ell, \sum_{\ell=1}^b |S_\ell|=t,\sum_{\ell=1}^b \sigma_J(S_\ell)=a\Big\}.$$
	Initialize $D(0,0,0)=0$, and all other entries as $-\infty$.
	Suppose the table for the first $b-1$ parts has been computed. 
	For $b=1,\dots,L$, the recurrence for the global table is
	\[
	D(b,t,a)=\max
	\Big\{D(b-1,t-q,a-u)+F_b(q,u): 0\le q\le \min\{t,m_b\},
	0\le u\le \min\{a,M_b\}\Big\},
	\]
	where infeasible entries are interpreted as $-\infty$.
	To justify the recurrence, suppose that in part $J_b$ we select a set
	$S_b\subseteq J_b$ such that $|S_b|=q$ and $\sigma_J(S_b)=u$.
	Since $J_b$ has $m_b$ elements and at most $M_b$
	pairs, we must have $0\le q\le m_b$ and $0\le u\le M_b$.
	For fixed $q$ and $u$, the best possible contribution of block $J_b$ is exactly $F_b(q,u)$. The previous $b-1$ blocks must then
	contribute cardinality $t-q$ and aggregate value $a-u$, which is captured by
	$D(b-1,t-q,a-u).$
	Taking the maximum over all feasible values of $q$ and $u$ gives the
	recurrence.
	
	After all parts have been processed, the value is $\max\{D(B,k,a)-\psi(a):a\in \{0,\dots,M\}\}.$
	Indeed, every feasible set $S\in\cS$ decomposes uniquely as $S=S_1\cup\cdots\cup S_B$, with $S_b=S\cap J_b$ for each $b\in [L]$.
	For each fixed aggregate value $a$, the table $D(B,k,a)$ gives the best
	possible value of the term $\sum_{i\in S}y_i$.
	Since $\mu$ is
	block-aggregable, subtracting $\psi(a)$ in the final maximization gives the correct objective
	value.
	
	Since $M=O(n^2)$, the running time for this dynamic program is polynomial in $n$.
	Combining this with the polynomial-time computation of the local tables
	$F_b$, the oracle problem can be solved in polynomial time.
\end{proof}

\noindent{\bf Applications.} As a direct consequence of \ref{thm:DPblock}, we can find apportionment rules that minimize the envy aggregation functions considered in Section~\ref{sec:envy-agg}.
More specifically, by using the function $\mu$ given by $\psi(x)=x$, we can solve \ref{form:minLP} to find the apportionment rule that minimizes total expected envy.
Then, by calling the function $\mu$ given by $\psi(x)=\ind{x>0}$, we can solve \ref{form:minLP} to find the apportionment rule that minimizes the probability of not being envy-free.
Finally, consider the case of single-floor instances.
As a corollary of both our Theorem~\ref{thm:rule-indep} and Theorem~\ref{thm:DPblock}, we obtain the following proposition.
\begin{proposition}\label{prop:single-floor-variance}
	In the single-floor setting, we can find in polynomial time the apportionment rule that minimizes the total envy variance.
\end{proposition}
\begin{proof}
	For any apportionment rule with distribution $S\mapsto p(S)$, we have that the variance of the total envy is equal to 
	\begin{align*}
		\text{Var}_{S\sim p}[\sigma_I(S)]&=\ee_{S\sim p}\Big[\sigma_I^2(S)\Big]-\textstyle\Big(\sum_{S\in \cS}\sigma_I(S)p(S)\Big)^2\\
		&=\sum_{S\in \cS}\sigma_I^2(S)p(S)-\textstyle\Big(\frac{1}{2}k(n-k) + \frac{1}{2}\sum_{t=1}^n r_t(n-2t+1)\Big)^2,
	\end{align*}
	where the second equality holds since the total expected envy is rule-independent for single-floor instances (Theorem~\ref{thm:rule-indep}).
	Therefore, to minimize the total envy variance, it suffices to minimize $\sum_{S\in \cS}\sigma_I^2(S)p(S)$.
	This objective is captured by the block-aggregable function with $\psi(x)=x^2$, and by calling \ref{form:minLP}, we can find the rule with minimum total envy variance.
\end{proof}

\begin{figure}[H]
	\centering
	\begin{subfigure}{0.6\textwidth}
		\centering
		\includegraphics[width=\linewidth]{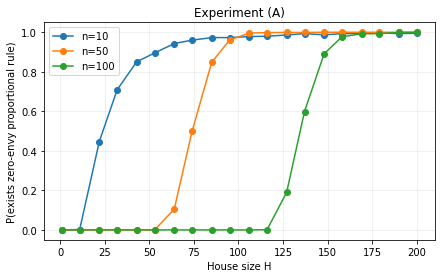}
		\caption{Probability of envy-free feasibility vs house size for various $n$.}
	\end{subfigure}
	\hfill
	\begin{subfigure}{0.6\textwidth}
		\centering
		\includegraphics[width=0.9\linewidth]{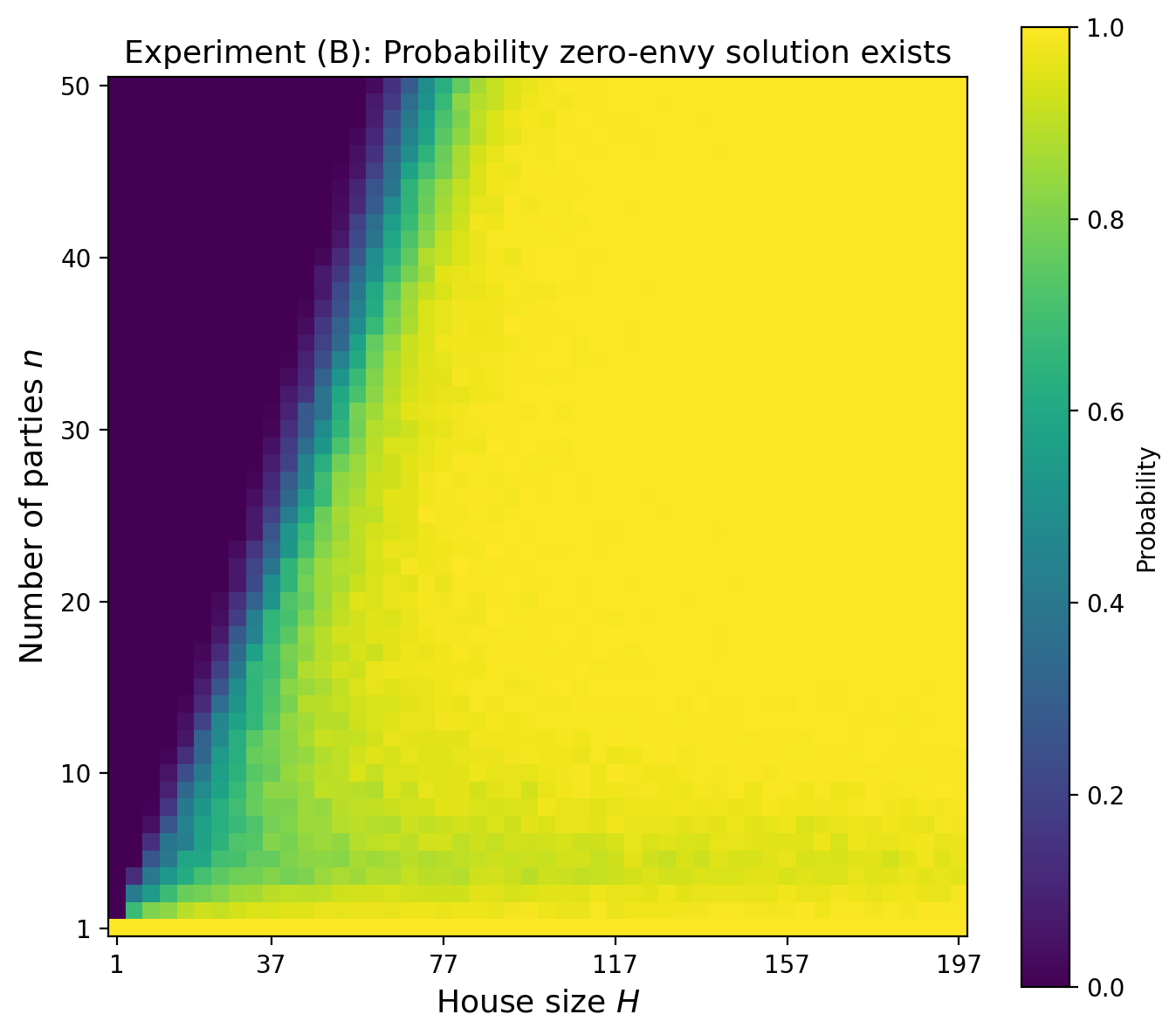}
		\caption{Heatmap of feasibility over $(n, H)$.}
	\end{subfigure}
	\hfill
	\begin{subfigure}{0.55\textwidth}
		\centering
		\includegraphics[width=\linewidth]{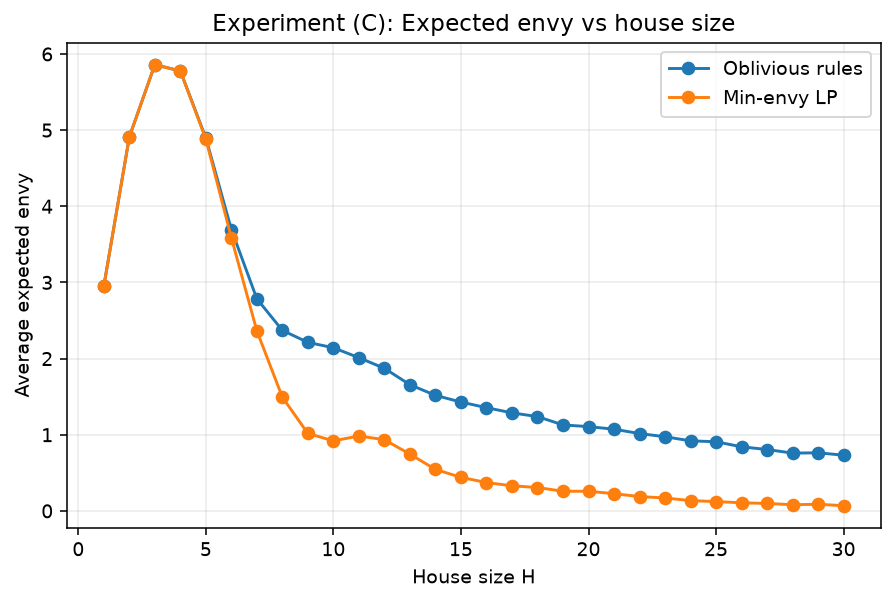}
		\caption{Expected envy: optimal vs.\txt{} oblivious.}
	\end{subfigure}
	
	\caption{Empirical behavior of envy-free feasibility and envy minimization. 
		Each plot is generated from synthetic instances with random vote vectors.}
	\label{fig:empirical-overview}
\end{figure}

\section{Empirical Results}\label{secEmpirical}

We complement the theoretical analysis with experiments designed to (i) measure how often an envy-free proportional randomized rule exists in practice, and (ii) compare the expected envy incurred by canonical apportionment rules to the minimum achievable envy.
Unless otherwise stated, each synthetic instance is generated by drawing a vote vector $v\in\mathbb{R}_{>0}^n$ i.i.d.\ from a continuous distribution (uniform on $[0,1]$), so ties occur with probability zero. Given $v$ and house size $H$, we compute quotas $q_i = H \cdot v_i/\sum_j v_j$ and decompose $q_i=f_i+r_i$. Parties are partitioned into equal-floor blocks $I_1,\dots,I_B$ by the floors $(f_i)$, and within each block we order by nondecreasing residues (breaking ties by index). Envy-free feasibility is decided by $\ref{form:EF}$ (the flow formulation from Section~\ref{subsecLP1}), solved with Gurobi via a direct \texttt{gurobipy} implementation~\cite{Gurobi}. The minimum expected total envy is computed for small $n$ by solving the explicit \ref{form:minLP} in Section~\ref{subsecMinEnvy} over all $k$-subsets.

Figure~\ref{fig:empirical-overview} summarizes the main empirical findings. Panel (a) reports the estimated probability that a zero-envy proportional randomized rule exists as a function of the house size $H$, for several values of $n$. Each point is computed over $1000$ independent random instances. The probability increases steadily with $H$, indicating that envy-free solutions become easier to realize as the number of seats grows relative to the number of parties. Intuitively, larger house sizes produce finer quota structures and more flexibility in matching marginal probabilities while avoiding within-block inversions. 

\begin{figure}[H]
	\centering
	\includegraphics[width=0.6\textwidth]{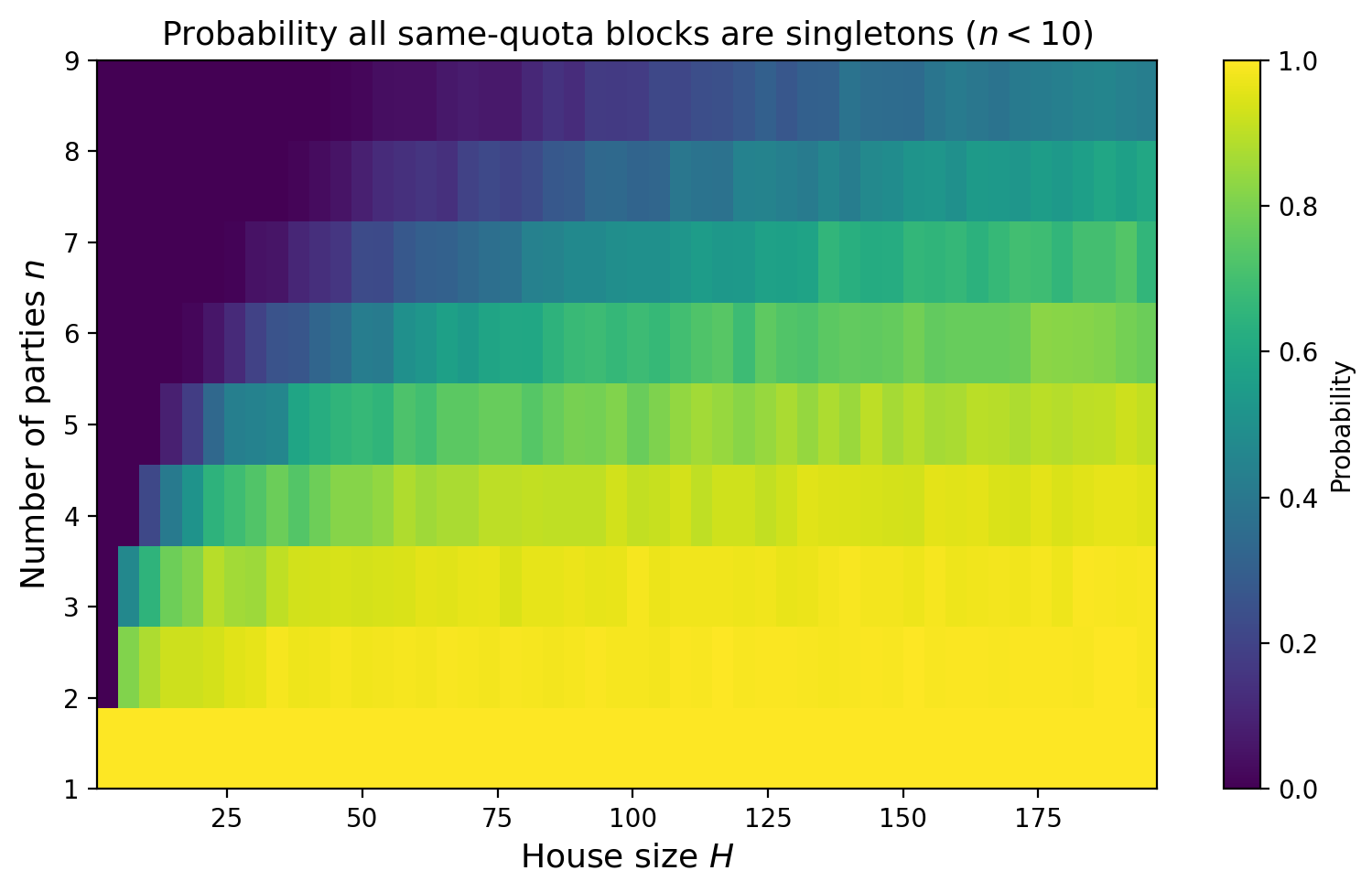}
	\caption{Probability that all same-quota-floor blocks are singletons,for $n < 10$ parties.}
	\label{fig:singleton}
\end{figure}

Panel (b) provides a complementary view by plotting feasibility over a two-dimensional grid of $(n,H)$. The heatmap shows that feasibility depends primarily on the ratio $H/n$ rather than on $n$ alone. For moderate values of $H$, envy-free solutions appear frequently across a wide range of problem sizes, suggesting that the phenomenon is not confined to small instances. A notable feature of Panel~(b) is that for small values of $n$ (e.g., $n \leq 10$), the feasibility probability does not reach~$1$ even for large house sizes~$H$. This is explained by the structure of same-floor blocks. Figure~\ref{fig:singleton} shows that for small~$n$, there is a persistent probability of encountering non-singleton blocks even at large~$H$: for instance, at $n=2$ and $H=197$, the probability that all blocks are singletons is approximately~$0.99$, closely matching the observed feasibility rate. Once $H/n$ is sufficiently large, the probability of non-singleton blocks vanishes and feasibility converges to~$1$. 

Panel (c) compares the expected envy achieved by different apportionment rules across $1000$ instances with $n = 10$ parties. We plot the minimum achievable envy (computed by solving \ref{form:minLP} exactly for small $n$) together with Grimmett's rule, the randomized-order variant of Grimmett, and Sampford rounding. Due to the symmetries of our distribution of votes, all of these oblivious rules consistently incur the same envy with a large number of trials. It is far larger envy than the optimum, sometimes by a substantial margin. These observations align with the theoretical result that oblivious rules can be quadratically suboptimal in the worst case, and suggest that exploiting floor-block structure is practically important even in moderately sized instances.

\subsection{U.S.\ House Apportionment over Census Decades}
We additionally tested envy-free feasibility on historical U.S.\ census populations and corresponding House sizes. The U.S.\ Constitution requires that each state receives at least one representative and that the District of Columbia receives none. To accommodate this constraint, we considered three alternative preprocessing methods, each guaranteeing at least one seat per state before applying the envy-free feasibility test.
We remark that in each case, after the preprocessing step, we get a new instance that is then used to feed our apportionment methods.

\paragraph{Option~1 (Baseline-1 proportional).}
Every state receives one deterministic seat. The remaining $H_{\mathrm{rem}} = H - n$ seats are then apportioned proportionally to population among all $n$ states, and we check envy-free feasibility of this residual problem via $\ref{form:EF}$.

\paragraph{Option~2 (Exclude tiny states).}
Every state again receives one baseline seat. We define \emph{tiny} states as those whose original proportional quota $q_i = p_i H / \sum_j p_j$ is less than~1. The residual $H_{\mathrm{rem}}$ seats are apportioned proportionally only among the non-tiny states, and $\ref{form:EF}$ is run on this restricted instance. Tiny states receive exactly one seat with certainty.

\paragraph{Option~3 (Quota-floor-1 reweighting).}
Rather than excluding tiny states, we construct effective vote weights $v_i^{\mathrm{eff}}$ such that the induced quotas satisfy $q_i \geq 1$ for all states. States whose proportional quota would fall below~1 are clamped to have effective quota exactly~1 (and hence residue~0, guaranteeing they never
receive a second seat), while the remaining quota mass is distributed proportionally among the other states. $\ref{form:EF}$ is then run on the full instance of $n$ states with house size~$H$.

\begin{figure}[H]
	\centering
	\begin{subfigure}{0.32\textwidth}
		\centering
		\includegraphics[width=\linewidth]{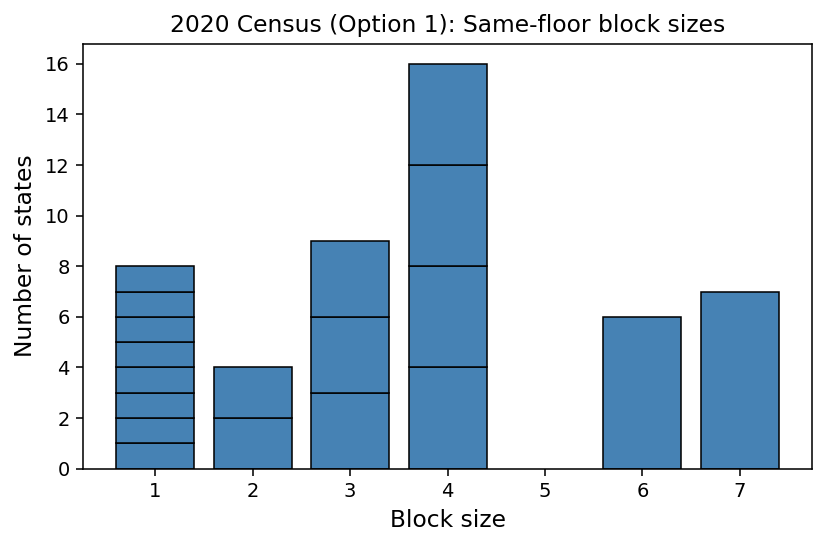}
		\caption{Option 1\\(Baseline-1 proportional)}
	\end{subfigure}
	\hfill 
	\begin{subfigure}{0.32\textwidth}
		\centering
		\includegraphics[width=\linewidth]{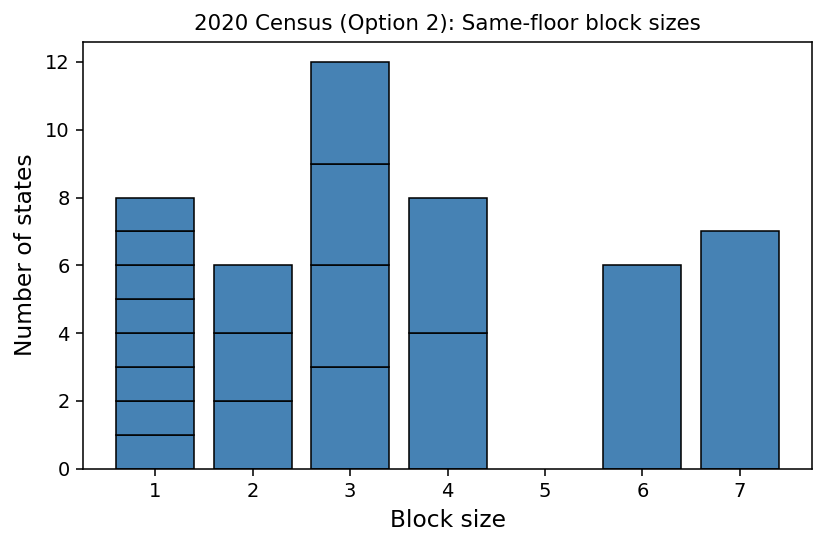}
		\caption{Option 2\\(Exclude tiny states)}
	\end{subfigure}
	\hfill 
	\begin{subfigure}{0.32\textwidth}
		\centering
		\includegraphics[width=\linewidth]{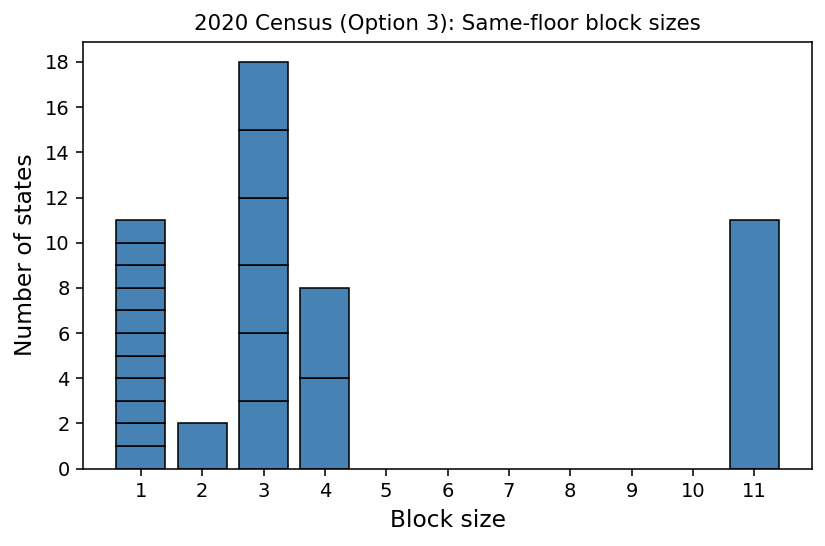}
		\caption{Option 3\\(Quota-floor-1 reweighting)}
	\end{subfigure}
	
	\caption{Same-floor block size distribution for the 2020 U.S.\ census under each preprocessing method. Each stacked segment represents one block; the height of each segment equals the block size.}
	\label{fig:blocks_2020}
\end{figure}

The differences between these three preprocessing methods are elaborated in Appendix~\ref{appThreeMethods}. Empirically we observed that an envy-free proportional rule exists for every tested decade under all three methods. This suggests that, at least for the historical U.S.\ instances, the floor-block structure admits sufficient flexibility to match the marginals while avoiding within-block inversions, regardless of how the constitutional minimum-seat constraint is handled. Figure~\ref{fig:blocks_2020} illustrates the same-floor block structure for the 2020 census under all three preprocessing methods. Across all options, block sizes remain moderate: most states fall into singleton or small blocks, and the largest blocks contain only a handful of states. The three methods produce similar but distinct distributions: Option~3, for instance, concentrates more states into fewer blocks due to the reweighting of tiny states, while Option~2 reduces the total number of states participating in the residual problem. Despite these differences, the overall structure remains favorable in every case. Since the LP resolves envy within each block independently, small blocks impose few binding constraints on the flow variables, which helps explain why envy-free feasibility holds across all decades and all three methods. As long as no single block grows large enough to create conflicting marginal requirements, feasibility is easily achieved.

\section{Conclusion}\label{secDiscussion}

In light of the contentious history surrounding apportionment, issues of fairness and paradoxes are paramount. To advance randomized apportionment from theory to practice, it is necessary to guard against outcomes that are likely to produce complaints, chief among which is envy. We have demonstrated that it is tractable to find envy-minimizing allocations at scale, and that the optimal level of envy is very often zero. These encouraging results suggest a clear path toward a wider adoption of randomized apportionment systems.

\appendix
\section{Notes on Preprocessing Methods for US Apportionment Data}\label{appThreeMethods}

\begin{table}[H]
	\centering
	\begin{tabular}{lrr}
		\toprule
		Pair & Avg $\ell_1$ & Avg $\ell_\infty$ \\
		\midrule
		Option 1 vs Option 2 & 3.2384 & 0.6383 \\
		Option 1 vs Option 3 & 25.2558 & 3.3754 \\
		Option 2 vs Option 3 & 23.0254 & 3.1946 \\
		\bottomrule
	\end{tabular}
	\caption{Average pairwise deviation in expected seats across all decades.}
	\label{tab:pairwise_dev}
\end{table}

The three methods differ in how they handle the tension between the constitutional minimum and proportionality. Fix~1 preserves proportional expectations across all states but may produce results with states with original quota below $1$ being assigned more than one seat. Option~2 eliminates tiny states from the randomized allocation entirely, improving feasibility at the cost of deviating from proportionality. Option~3 occupies a middle ground: all states participate in the LP, but tiny states are effectively frozen at one seat through reweighting. Table~\ref{tab:pairwise_dev} reports the average pairwise deviation in expected seat allocations across all census decades, measured
in both $\ell_1$ and $\ell_\infty$ norm.

\bibliographystyle{siamplain}
\bibliography{bibliography}

@book{grotschel2012geometric,
  title={Geometric algorithms and combinatorial optimization},
  author={Gr{\"o}tschel, Martin and Lov{\'a}sz, L{\'a}szl{\'o} and Schrijver, Alexander},
  volume={2},
  year={2012},
  publisher={Springer Science \& Business Media}
}

@article{cembrano2022multidimensional,
  title={Multidimensional political apportionment},
  author={Cembrano, Javier and Correa, Jos{\'e} and Verdugo, Victor},
  journal={Proceedings of the National Academy of Sciences},
  volume={119},
  number={15},
  pages={e2109305119},
  year={2022}
}

@article{magnanti1993network,
  title={Network flows},
  author={Magnanti, Thomas L and Orlin, James B},
  journal={PHI, Englewood Cliffs NJ},
  year={1993}
}

@article{BadNoApportionment,
  author  = {Richard Boeckel },
  title   = {Apportionment of Representatives in Congress},
  journal = {CQ Researcher},
  year    = {1927},
  url     = {https://cqpress.sagepub.com/cqresearcher/report/apportionment-representatives-congress-cqresrre1927120600},
  note    = {Accessed: 2026-02-09}
}

@book{Pukelsheim17,
  title = {Proportional Representation: Apportionment Methods and Their Applications},
  shorttitle = {Proportional Representation},
  author = {Pukelsheim, Friedrich},
  year = {2017},
  edition = {2nd edition},
  address = {{New York, NY}},
  publisher = {{Springer}}
}

@article{CembranoOnlineApportionment,
  title={Online Proportional Apportionment},
  author={Cembrano, Javier and Correa, Jose and Griesbach, Svenja M and Verdugo, Victor},
  journal={Mathematical Programming},
  year={2026}
}

@article{Cembrano2025Proportionality,
  author  = {Javier Cembrano and José Correa and Gonzalo Díaz and Victor Verdugo},
  title   = {Proportionality in multiple dimensions to design electoral systems},
  journal = {Social Choice and Welfare},
  year    = {2025},
  publisher = {Springer},
  pages   = {1--35}
}

@book{BY01,
  title = {Fair Representation: Meeting the Ideal of One Man, One Vote},
  shorttitle = {Fair Representation},
  author = {Balinski, Michel L. and Young, H. Peyton},
  year = {2001},
  edition = {2nd},
  publisher = {{Brookings Institution Press}}
}

@article{BY74,
  title={A new method for congressional apportionment},
  author={Balinski, Michel L. and Young, H. Peyton},
  journal={Proceedings of the National Academy of Sciences},
  volume={71},
  number={11},
  pages={4602--4606},
  year={1974}
}

@article{BY75,
  title = {The Quota Method of Apportionment},
  author = {Balinski, Michel L. and Young, H. Peyton},
  year = {1975},
  journal = {The American Mathematical Monthly},
  volume = {82},
  number = {7},
  pages = {701--730}
}

@article{BY77,
  title = {On {{Huntington Methods}} of {{Apportionment}}},
  author = {Balinski, Michel L. and Young, H. Peyton},
  year = {1977},
    journal = {SIAM Journal on Applied Mathematics},
  volume = {33},
  number = {4},
  pages = {607--618}
}

@article{BY79,
  title = {Quotatone Apportionment Methods},
  author = {Balinski, Michel L. and Young, H. Peyton},
  year = {1979},
  journal = {Mathematics of Operations Research},
  volume = {4},
  number = {1},
  pages = {31--38},
  publisher = {{INFORMS}}
}

@article{BY79a,
  title = {Criteria for {{Proportional Representation}}},
  author = {Balinski, M. L. and Young, H. P.},
  year = {1979},
    journal = {Operations Research},
  volume = {27},
  number = {1},
  pages = {80--95}
}

@article{CembranoCombInsights,
  title={New combinatorial insights for monotone apportionment},
  author={Cembrano, Javier and Correa, Jos{\'e} and Schmidt-Kraepelin, Ulrike and Tsigonias-Dimitriadis, Alexandros and Verdugo, Victor},
  journal={Mathematics of Operations Research},
  year={2025},
  publisher={INFORMS}
}

@article{GoltzHouse,
author = {G\"{o}lz, Paul and Peters, Dominik and Procaccia, Ariel D.},
title = {In This Apportionment Lottery, the House Always Wins},
journal = {Operations Research},
volume = {74},
number = {1},
pages = {390-407},
year = {2026}
}

@article{GaffkeDivisor,
title = {Divisor methods for proportional representation systems: An optimization approach to vector and matrix apportionment problems},
journal = {Mathematical Social Sciences},
volume = {56},
number = {2},
pages = {166-184},
year = {2008},
author = {Norbert Gaffke and Friedrich Pukelsheim}
}

@article{Grimmett,
 author = {Geoffrey Grimmett},
 journal = {The American Mathematical Monthly},
 number = {4},
 pages = {299--307},
 publisher = {Mathematical Association of America},
 title = {Stochastic Apportionment},
 volume = {111},
 year = {2004}
}

@article{Sampford,
  title={On sampling without replacement with unequal probabilities of selection},
  author={Sampford, Michael R},
  journal={Biometrika},
  volume={54},
  number={3-4},
  pages={499--513},
  year={1967}
}

@article{MonotoneRandomizedApportionment,
author = {Correa, Jos\'{e} and G\"{o}lz, Paul and Schmidt-Kraepelin, Ulrike and Tucker-Foltz, Jamie and Verdugo, Victor},
title = {Monotone Randomized Apportionment},
year = {2026},
journal ={Operations Research},
publisher ={INFORMS}
}

@inproceedings{RoteMatrix,
author = {Rote, G\"{u}nter and Zachariasen, Martin},
title = {Matrix scaling by network flow},
year = {2007},
booktitle = {SODA},
pages = {848–854},
numpages = {7}
}

@misc{Gurobi,
  author = {{Gurobi Optimization, LLC}},
  title = {{Gurobi Optimizer Reference Manual}},
  year = 2021,
  url = "https://www.gurobi.com"
}

@article{Madow49,
	title = {On the {{Theory}} of {{Systematic Sampling}}, {{II}}},
	author = {Madow, William G.},
	year = {1949},
	journal = {The Annals of Mathematical Statistics},
	volume = {20},
	number = {3},
	pages = {333--354}
}

@article{BestOfBothWorlds,
	author       = {Haris Aziz and
	Rupert Freeman and
	Nisarg Shah and
	Rohit Vaish},
	title        = {Best of Both Worlds: Ex Ante and Ex Post Fairness in Resource Allocation},
	journal      = {Operations Research},
	volume       = {72},
	number       = {4},
	pages        = {1674--1688},
	year         = {2024}
}

@inproceedings{CitySampling,
	author       = {Paul G{\"{o}}lz and
	Jan Maly and
	Ulrike Schmidt{-}Kraepelin and
	Markus Utke and
	Philipp C. Verpoort},
	title        = {City Sampling for Citizens' Assemblies},
	booktitle    = {AAAI},
	pages        = {16971--16979},
	year         = {2026}
}

@article{Blotto,
	title = {The multiplayer Colonel Blotto game},
	journal = {Games and Economic Behavior},
	volume = {129},
	pages = {15-31},
	year = {2021},
	author = {Enric Boix-Adserà and Benjamin L. Edelman and Siddhartha Jayanti}
}

@book{TilleBookAlgorithms,
	author = {Tillé, Yves. author.},
	year = {2006},
	publisher = {Springer},
	title = {Sampling Algorithms},
}

@article{AzizPeerSelection,
	title={Strategyproof peer selection using randomization, partitioning, and apportionment},
	author={Aziz, Haris and Lev, Omer and Mattei, Nicholas and Rosenschein, Jeffrey S and Walsh, Toby},
	journal={Artificial Intelligence},
	volume={275},
	pages={295--309},
	year={2019}
}

@article{Gandhi,
	title = {Dependent Rounding and Its Applications to Approximation Algorithms},
	author = {Gandhi, Rajiv and Khuller, Samir and Parthasarathy, Srinivasan and Srinivasan, Aravind},
	year = {2006},
	journal = {Journal of the ACM},
	volume = {53},
	number = {3},
	pages = {324--360}
}

@misc{AffirmativeAction,
	author = {Evren, Haydar and Khanna, Manshu},
	title  = {Affirmative Action's Cumulative Fractional Assignments},
	year   = {2021}
}

@book{CitizensAssemblies,
	title = {Innovative {{Citizen Participation}} and {{New Democratic Institutions}}: {{Catching}} the {{Deliberative Wave}}},
	shorttitle = {Innovative {{Citizen Participation}} and {{New Democratic Institutions}}},
	author = {{OECD}},
	year = {2020},
	month = jun,
	publisher = {Organisation for Economic Co-operation and Development},
	doi = {10.1787/339306da-en},
	langid = {english},
	shortauthor = {OECD}
}

@article{ProbabilisticSerial,
	title = {A New Solution to the Random Assignment Problem},
	journal = {Journal of Economic Theory},
	volume = {100},
	number = {2},
	pages = {295-328},
	year = {2001},
	author = {Anna Bogomolnaia and Hervé Moulin}
}

@inproceedings{IndivisibleGoodsSurvey,
	author       = {Georgios Amanatidis and
	Georgios Birmpas and
	Aris Filos{-}Ratsikas and
	Alexandros A. Voudouris},
	title        = {Fair Division of Indivisible Goods: {A} Survey},
	booktitle    = {IJCAI},
	pages        = {5385--5393},
	year         = {2022}
}

@book{BrewerHanif,
	title = {An {{Introduction}} to {{Sampling}} with {{Unequal Probabilities}}},
	author = {Brewer, K. R. W. and Hanif, Muhammad},
	year = {1983},
	volume = {15},
	address = {{New York, NY}},
	publisher = {{Springer}}
}
\end{document}